\documentclass[preprintnumbers,superscriptaddress,endnote,nofootinbib,aps,prd,floatfix]{revtex4}

\usepackage[utf8x]{inputenc}     % if unicode errors are coming
\usepackage{enumerate}
\usepackage{epsfig} 
\usepackage{float} 
\usepackage[caption = false]{subfig}
\usepackage{amsmath,amssymb}
\usepackage{wasysym} 
\usepackage{mathrsfs} 
\usepackage{amsfonts} 
\usepackage{amsbsy} 
\usepackage{amscd} 
\usepackage{pstricks} 
\usepackage{multirow} 
\usepackage{tikz}
\usepackage{color}
\usepackage{slashed}
\usepackage{array}
\usepackage{tablefootnote}
\usepackage{csvsimple} 
\usetikzlibrary{arrows,positioning,shapes.geometric} 
\usepackage[compat=1.1.0]{tikz-feynman}         
\usepackage{slashed}
\usepackage[font=small,labelfont=bf,
justification=raggedright,
format=plain]{caption}
\usepackage{multirow}
\usepackage{tabularx}
\usepackage{soul}  
\usepackage{listings} 
\usepackage{hyperref}
\hypersetup{colorlinks=true, citecolor=blue, urlcolor=blue, linkcolor=blue}
\usepackage{color}
\newcommand{\ket}[1]{|#1\rangle}
\newcommand{\bra}[1]{\langle#1|}
\newcommand{\braket}[2]{\langle#1|#2\rangle}

\newcommand{\trace}[2]{\text{Tr}_{#1}\;\left \{ #2 \right \} }
\usepackage{amsthm}

\begin{document}
%%%%%%%%%%%%%%%%%%%%%%%%%%%%%%%%%%%%%%%%%%%%%%%%%% 
\title{Anomaly detection in high-energy physics using a quantum autoencoder}

%\author[a,b]{Vishal~S.~Ngairangbam,}
%\author[c,d]{Michael~Spannowsky,} 
%\author[e]{and Michihisa Takeuchi}
%\affiliation[a]{Theoretical Physics Division, Physical Research Laboratory,\\ Shree Pannalal Patel Marg, Ahmedabad - 380009, Gujarat, India}
%\affiliation[b]{Discipline of Physics, Indian Institute of Technology, Palaj,\\ Gandhinagar - 382424, Gujarat, India}
%\affiliation[c]{Institute for Particle Physics Phenomenology, Durham University,\\ Durham DH1 3LE, United Kingdom}
%\affiliation[d]{Department of Physics, Durham University,\\ Durham DH1 3LE, United Kingdom}
%\affiliation[e]{Department of Physics, Osaka University, Osaka 560-0043, Japan}
%\emailAdd{vishalng@prl.res.in}
%
%\emailAdd{michael.spannowsky@durham.ac.uk}
%\emailAdd{m.takeuchi@het.phys.sci.osaka-u.ac.jp}
\begin{abstract}
	The lack of evidence for new interactions and particles at the Large Hadron Collider has motivated the high-energy physics community to explore model-agnostic data-analysis approaches to search for new physics. Autoencoders are unsupervised machine learning models based on artificial neural networks, capable of learning background distributions. We study quantum autoencoders based on variational quantum circuits for the problem of anomaly detection at the LHC. For a QCD $t\bar{t}$ background and resonant heavy Higgs signals, we find that a simple quantum autoencoder outperforms classical autoencoders for the same inputs and trains very efficiently. Moreover, this performance is reproducible on present quantum devices. This shows that quantum autoencoders are good candidates for analysing high-energy physics data in future LHC runs.   
\end{abstract}
\preprint{\today, OU-HET-1125, IPPP/21/54}

%\keywords{Large Hadron Collider, Anomaly Detection, Quantum Machine Learning, Quantum Autoencoder}
%%%%%%%%%%%%%%%%%%%%%%%%%%%%%%%%%%%%%%%%%%%%%%%%%% 
%==========================================================================
\author{Vishal S. Ngairangbam}\email{vishalng@prl.res.in}
\affiliation{Physical Research Laboratory, Ahmedabad - 380009, Gujarat, India\\[0.1cm]}
\affiliation{Discipline of Physics, Indian Institute of Technology, Palaj,\\ Gandhinagar - 382424, Gujarat, India}

\author{Michael~Spannowsky} \email{michael.spannowsky@durham.ac.uk}
\affiliation{Institute for Particle Physics Phenomenology, Durham University, Durham DH1 3LE, United Kingdom\\[0.1cm]}

\author{Michihisa Takeuchi} \email{m.takeuchi@het.phys.sci.osaka-u.ac.jp}
\affiliation{Department of Physics, Osaka University, Osaka 560-0043, Japan\\[0.1cm]}

\maketitle
\flushbottom
%\tableofcontents
\section{Introduction}

In the absence of a confirmed new physics signal and in the presence of a plethora of new physics scenarios that could hide in the copiously produced LHC collision events, unbiased event reconstruction and classification methods\cite{ATLAS:2020iwa,CMS:2020zjg,Kasieczka:2021xcg,Aarrestad:2021oeb} have become a major research focus of the high-energy physics community. Unsupervised machine learning models~\cite{Fraser:2021lxm,Ostdiek:2021bem,Andreassen:2018apy,Dorigo:2021iyy}, popularly used as anomaly-detection methods~\cite{DAgnolo:2018cun,Aguilar-Saavedra:2017rzt,Blance:2019ibf,Mikuni:2021nwn,Govorkova:2021utb,Hajer:2018kqm,Roy:2019jae}, are trained on Standard Model processes and should indicate if a collision event is irreconcilable with the kinematic features of events predicted by the Standard Model. 

One of the most popular neural network-based approach are autoencoders \cite{autoenc}. Autoencoders consist of an encoder step that compresses the input features into a latent representation with reduced dimensionality. Subsequently, the latent representation is decoded into an output of the same dimensionality as the input feature space. The entire network is then trained to minimise the reconstruction error. The latent space acts as an information bottleneck, and its dimension is a hyperparameter of the network. The assumption is that the minimal dimension of the latent space for which the input features can still be reconstructed corresponds to the intrinsic dimension of the input data, here Standard Model induced background processes. However, the trained autoencoder would poorly reconstruct any unknown new-physics process with a higher intrinsic dimension.  If the signal is kinematically sufficiently different from the background samples, the loss or reconstruction error will be larger for signal than for background events. Such autoencoders can be augmented with convolutional neural networks \cite{Heimel:2018mkt,Farina:2018fyg}, graph neural networks \cite{Atkinson:2021nlt,Tsan:2021brw} or recurrent neural networks \cite{davino2017autoencoder,Marchi115520174694860-zb} on its outset, making it a very flexible anomaly detection method for a vast number of use cases. 
 
With the advent of widely available noisy intermediate-scale quantum computers (NISQ) \cite{Preskill2018quantumcomputingin} the interest in quantum algorithms applied to high-energy physics problems has spurred. Today's quantum computers have a respectable quantum volume and can perform highly non-trivial computations. This technical development has resulted in a community-wide effort~\cite{Feynman1982,Georgescu:2013oza} exploring the applications of quantum computers for studying quantum physics in general and in particular, the application to challenges in the theoretical description of particle physics. Some recent studies in the direction of LHC physics include evaluating Feynman loop integrals~\cite{Ramirez-Uribe:2021ubp}, simulating parton showers~\cite{Williams:2021lvr} and structure~\cite{Li:2021kcs}, quantum algorithm for evaluating helicity amplitudes~\cite{Bepari:2020xqi}, and simulating quantum field theories~\cite{Jordan:2014tma,Preskill:2018fag,Bauer:2019qxa,Abel:2020ebj,Abel:2020qzm,Davoudi:2021ney}. An interesting application of quantum computers is the nascent field of quantum machine learning--leveraging the power of quantum devices for machine learning tasks, with the capability of classical\footnote{By classical, we mean any machine learning algorithm that leverages only discrete bit computations, while by quantum, we imply a computation that uses the properties of quantum mechanics and qubits, even if they are simulated on classical hardware.} machine learning algorithms for various applications at the LHC already recognised, it is only natural to explore whether quantum machine learning (QML) can improve the classical algorithms~\cite{Mott2017,Blance:2020ktp,Wu:2020cye,Blance2021,Abel:2021fpn,Wu:2021xsj,Chen:2021ouz,Terashi:2020wfi}.

This work explores the feasibility and potential advantages of using quantum autoencoders (QAE) for anomaly detection. Most quantum algorithms consist of a quantum state, encoded through qubits, which evolves through the application of a unitary operator. The necessary compression and expansion of data in the encoding and decoding steps are manifestly non-unitary, which has to be addressed by the QAE using entanglement operations and reference states which disallow information to flow from the encoder to the decoder. To this end, a QAE should, in principle, be able to perform tasks ordinarily accomplished by a classical autoencoder (CAE) based on deep neural networks (DNN). The ability of DNNs are known to scale with data \cite{Sarker2021}, and large datasets are necessary to bring out their better performance over other machine-learning algorithms. Interestingly, we find that a quantum autoencoder, augmented using quantum gradient descent \cite{Stokes2020quantumnatural,Blance2021} for its training, is much less dependent on the number of training samples and reaches optimal reconstruction performance with minuscule training datasets. Since the use of quantum gradient descent is a relatively new way of improving the convergence speed and reliability of the quantum network training, we provide a detailed introduction in Appendix \ref{app:qgd}. Moreover, compared to CAEs, which use the same input variables as the QAE, QAEs have better anomaly detection capabilities for the two benchmark processes we use in our study. This better performance is particularly interesting as the CAE has $\mathcal{O}(1000)$ parameters compared to just $\mathcal{O}(10)$ for the QAE. The study indicates the possibility to study quantum latent representations of high-energy collisions, in analogy to classical autoencoders~\cite{Dillon:2021nxw,Bortolato:2021zic,Atkinson:2021nlt,Dillon:2019cqt}.  Our results indicate that quantum autoencoders could be advantageous in anomaly detection tasks in the NISQ era.

The rest of the paper is organised as follows. In section~\ref{sec:cae}, we present an introduction to classical autoencoders based on deep neural networks. We then describe the basic ideas of quantum machine learning and a quantum autoencoder in section~\ref{sec:qae}. The details of the data simulation, network architecture, and training are described in section~\ref{sec:an_setup}. We present the performance of a quantum autoencoder compared to a classical autoencoder in section~\ref{sec:results}. We conclude in section~\ref{sec:conc}. 

\section{Classical Autoencoders}
\label{sec:cae} 
\begin{figure*}[th!]
	{\centering 
		\includegraphics[scale=.85]{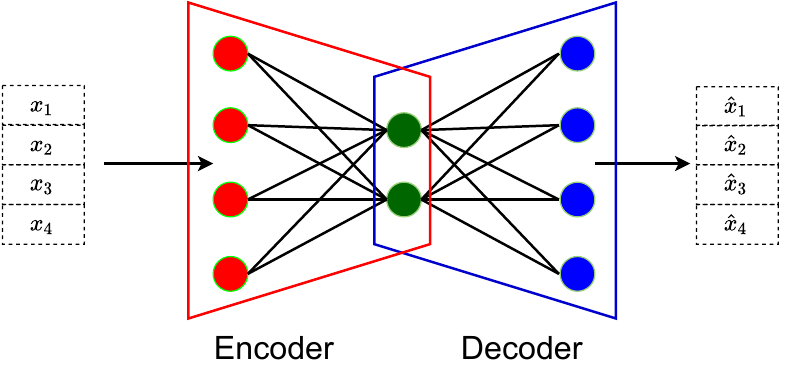}
		%\hspace{1cm}
		\includegraphics[scale=0.358]{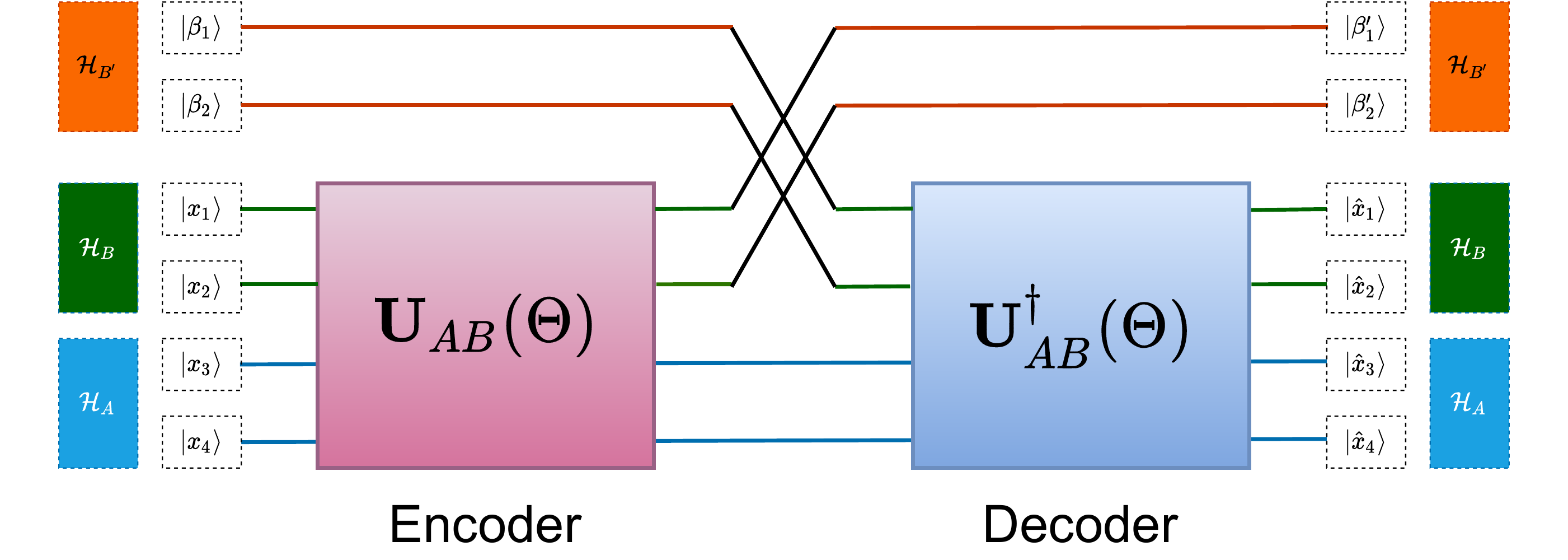}
	}
	\caption{Schematic representation of a simple dense classical autoencoder (left) and a quantum autoencoder (right) for a four dimensional input space and a two dimensional latent space. To induce an information bottleneck in quantum unitary evolutions, we throw away states $\ket{\beta'_i}$ (trash states) at the encoder output (green lines), which are replaced by reference states $\ket{\beta_i}$ (shown in orange lines ), containing no information of the input $\ket{x_j}$. The mechanism can be better understood by dividing the Hilbert space of the complete system into three parts: $\mathcal{H}_A$ the subspace formed by the qubits that are fed to the decoder, $\mathcal{H}_B$ the subspace of the qubits that are discarded after encoding, and $\mathcal{H}_{B'}$ the subspace where a fixed reference state (initialised as $\ket{0}^{\otimes \dim{\mathcal{H_B}}})$ unacted by the encoder is fed to the decoder. SWAP gates can achieve the exchange of states denoted by black lines. 
		}
	\label{fig:cae_qae} 
	
\end{figure*}
Autoencoders are neural networks utilised in various applications of unsupervised learning. They learn to map input vectors $\mathbf{x}$ to a \emph{compressed} latent vector $\mathbf{z}$ via an encoder. This latent vector feeds into a decoder that reconstructs the inputs. Denoting the encoder and decoder networks as $\mathbf{E}(\Theta_E,\mathbf{x})$ and $\mathbf{D}(\Theta_D,\mathbf{z})$ with $\Theta_E$ and $\Theta_D$ denoting the learnable parameters of the respective network, we have
\begin{equation}
\label{eq:cls_autoenc}
\mathbf{z}=\mathbf{E}(\Theta_E,\mathbf{x})\quad,\quad \hat{\mathbf{x}}=\mathbf{D}(\Theta_D,\mathbf{z})\quad,
\end{equation}
where $\hat{\mathbf{x}}$ denote the reconstructed output vector. The whole network is trained via gradient descent to reduce a faithful distance $L$, between the reconstructed output $\hat{\mathbf{x}}$  and the input vector $\mathbf{x}$. For instance $L$ can be the root-mean-square-error (RMSE),
\begin{equation}
\label{eq:rmse}
L(\mathbf{x},\hat{\mathbf{x}})=\sqrt{\frac{\sum_{i=1}^{i=n}(\hat{x}^i-x^i)^2}{n}}\quad,
\end{equation} where $\hat{x}^i$ and $x^i$ are the $i^{th}$ component of the reconstructed and input vectors respectively, and $n$ is their dimension.  A faithful encoding should have an optimal latent dimension $k<n$, with $k$ being the intrinsic dimension of the data set. This \emph{dimensionality reduction} is crucial in many applications of autoencoders, which otherwise learns trivial mappings to reconstruct the output vectors $\hat{x}$. Unsupervised learning deals with learning probability distributions, and properly trained autoencoders are excellent for many applications. A dense CAE for a four feature input and two-dimensional latent space is shown in figure~\ref{fig:cae_qae}. The encoder and the decoders are also enclosed in red and blue boxes, respectively.     

One popular usage of autoencoders in collider physics is anomaly detection.  In various scenarios at the LHC, the background processes' contributions are orders of magnitude larger than most viable signals. However, a plethora of possible signal scenarios exist that could be realised in nature, making it unlikely that the signal-specific reconstruction techniques of supervised learning methods comprehensively cover all possible scenarios. This motivates unsupervised anomaly detection techniques, wherein a statistical model learns the probability distribution of the background to classify any data not belonging to it as anomalous (signal) data. Using an autoencoder as an anomaly detector, we train it to reconstruct the background data faithfully. Many signals have a higher intrinsic dimension\footnote{It has been found in ref~\cite{Finke:2021sdf} that Convolutional Autoencoders cannot detect signals of lower intrinsic dimensions. While Quantum Autoencoders could alleviate this issue, we do not study their properties for lower dimensional signals. The study aims to validate their workings on similar scenarios where classical autoencoders work. } than background data due to their increased complexity. Hence, they incur higher reconstruction losses. Thus, the loss function can be used as a discriminant to look for anomalous events.  

\section{Quantum autoencoders}

\label{sec:qae}   
Quantum machine learning broadly deals with extending classical machine learning problems to the quantum domain with variational quantum  circuits~\cite{var_qml}.  We can divide these circuits into three blocks: a state preparation that encodes classical inputs into quantum states, a unitary evolution circuit that evolves the input states, and a measurement and post-processing part that measures the evolved state and processing the obtained observables further. For this discussion, we will always work in the computational basis with the basis vectors $\{\ket{0},\ket{1}\}$ denoting the eigen states of the Pauli Z operator $\hat{\sigma}_z$ for each qubit.

There are many examples of state preparation in literature~\cite{robust_enc}, which has their own merits in various applications. We prepare the states using {\it angle encoding}, which encodes real-valued observables $\phi_j$ as rotation angles along the $x$-axis of the Bloch sphere
\begin{equation}
\label{eq:state_prep}
\ket{\Phi} = \bigotimes_{i=1}^n R_x(\phi_j)\;\ket{0}=\bigotimes_{j=1}^n \;\left(\cos\frac{\phi_j}{2}\;\ket{0}-i\sin\frac{\phi_j}{2}\ket{1}\right)~,
\end{equation} where $R_x=e^{-i\frac{\phi_j}{2}\hat{\sigma}_x}$ denote the rotation matrix. The number of qubits required $n$, is same as the dimensions of the input vector. A parametrised unitary circuit $\mathbf{U}(\Theta)$, with $\Theta$ denoting the set of parameters, evolves the prepared state $\ket{\Phi}$ to a final state $\ket{\Psi}$,
\begin{equation}
\label{eq:uni_eve}
\ket{\Psi}=\mathbf{U}(\Theta)\;\ket{\Phi}~.
\end{equation} The final measurement step involves the measurement of an observable on the final state $\ket{\Psi}$.  Since measurements in quantum mechanics are inherently probabilistic, we measure multiple times (called shots) to get an accurate result. In order to do that, we need quantum hardware that can prepare a large number of pure identical input states $\ket{\Phi}$ for each data point. 

After defining a cost function, the parameters $\Theta$ can be trained and updated using an optimisation method.
To better capture the geometry of the underlying Hilbert space and to achieve a faster training of the quantum network,\footnote{See \cite{Blance2021} for a brief presentation of the Fubini-Study metric and a comparison of natural and quantum gradient descent for the training of classical and quantum networks. It was shown that quantum gradient descent improves the training of a variational quantum circuit significantly.} we will use quantum gradient descent~\cite{Stokes2020quantumnatural}, where the direction of steepest descent is evaluated according to the Fubini-Study metric~\cite{fubini,Study1905}. The general idea is to make the optimisation procedure aware of the weight space's underlying quantum geometry, which improves the speed and reliability of finding the global minimum of the loss function. A brief outline of quantum gradient descent is given in Appendix~\ref{app:qgd}.

While we have not discussed the specific form of the parametrised unitary operation $\mathbf{U}(\Theta)$, it is important to note that one of the major advantages of quantum computation is due to its ability to produce \emph{entangled states}, a phenomenon absent in devices based on classical bits. The prepared input state is separable into the component qubits, and a product of unitaries acting on single-qubit states will not entangle the subsystems. The CNOT gate is a standard two-qubit gate, which will be used in our circuit to entangle the subsystems. 
\subsection{Quantum autoencoders on variational circuits}
Quantum autoencoders based on variational circuit models have been proposed for quantum data compression~\cite{romero2017quantum}. In our work, we want to learn the parameters of such a network to compress the background data efficiently. Along the same principles as anomaly detection on classical autoencoders, we expect that the compression and subsequent reconstruction will work poorly on data with different characteristics to the background. 

A quantum autoencoder, in analogy to the classical autoencoders has an encoder circuit which evolves the input state $\ket{\Phi}$ to a latent state $\ket{\chi}$ via a unitary transformation $\mathbf{U}(\Theta)$, and then reconstructs the input state, via its hermitian conjugate $\ket{\Phi}=\mathbf{U}^\dagger(\Theta)\ket{\chi}$. However, note that since unitary transformations are probability conserving and act on spaces having identical dimensions, there is no data compression in such a setup.  In order to have data compression, some qubits at the initial encoding $\ket{\chi}$ are discarded and replaced by freshly prepared reference states. Such a setup for a four feature input and two dimensional latent space if shown in figure~\ref{fig:cae_qae}. The unitary operators output identical number of qubits, however at the encoder step, two of its outputs (shown by green lines) are replaced by freshly prepared reference states (shown in orange lines), devoid of any information of the input states. We describe the basics of quantum autoencoding in the following, mainly based on the discussion of quantum autoencoders for data compression from ref.~\cite{romero2017quantum}. Quantum anomaly detection of simulated quantum states has been investigated in ref~\cite{Kottmann:2021usw}. To the best of our knowledge, our study is the first to explore anomaly detection of classical inputs via a quantum autoencoder. The main difference between existing studies and ours is that the input states for the former are inherently quantum mechanical. In contrast, the choice of input embedding of the classical numbers in our case determines the nature of the quantum state. We will use angular encoding, where the quantum states are separable into the constituent qubits. We will, however, be extensively using CNOT gates in the unitary evolution which will entangle the different qubits.

Let us denote the Hilbert space containing the input states by $\mathcal{H}$. For describing a quantum autoencoder, it is convenient to expand $\mathcal{H}$ as the product of three subspaces, 
\begin{equation}
\label{eq:hilb_sub}
\mathcal{H}=\mathcal{H}_A\otimes\mathcal{H}_B\otimes\mathcal{H}_{B'}~,
\end{equation}  with subspace $\mathcal{H}_A$ denoting the space of qubits fed into the decoder from the encoder, and $\mathcal{H}_B$ denoting the space corresponding to the ones that are re-initialised, and $\mathcal{H}_{B'}$ denoting the Hilbert space containing the reference state.  
In the following, we will denote states belonging to any subspace with suffixes while the full set will have no suffix. For example, $\ket{a}_{AB}\in \mathcal{H}_A\otimes \mathcal{H}_B$, $\ket{\kappa}\in \mathcal{H}$, $\ket{b}_{B'}\in \mathcal{H}_{B'}$ etc. We will use the same convention for operators acting on the various subspaces. 

Since we entangle the separable input qubits in the subspaces $\mathcal{H}_A\otimes\mathcal{H}_B$ via $\mathbf{U}_{AB}(\Theta)$, the latent state $\ket{\chi}_{AB}\in \mathcal{H}_A\otimes\mathcal{H}_B$ , in general, is not seperable. The input of the larger composite system including the reference state is $\ket{\Phi}_{AB}\otimes\ket{\beta}_{B'}$, with $\ket{\beta}_{B'}$ denoting a freshly prepared \emph{reference state} (initialised as $\ket{0}^{\otimes \dim{\mathcal{H_B'}}}$) not acted on by the unitary $\mathbf{U}_{AB}$. The process of encoding can be therefore written as,
\begin{equation}
\label{eq:enc_quantum}
\ket{\chi}_{AB}\otimes\ket{\beta}_{B'}=(\mathbf{U}_{AB}(\Theta)\otimes\mathbf{I}_{B'})\;\ket{\Phi}_{AB}\otimes\ket{\beta}_{B'}\quad,
\end{equation}where $\mathbf{I}_{B'}$ denotes the identity operator on $\mathcal{H}_{B'}$. Explicitly, the dimensions of the subspaces $\mathcal{H}_A$, $\mathcal{H}_B$, and $\mathcal{H}_{B'}$ are $2^{N_{lat}}$, $2^{N_{trash}}$, and $2^{N_{trash}}$, respectively, where $N_{lat}$ is the number of qubits passed to the decoder directly from the encoder, while $N_{trash}$ are the ones that are discarded.
Swapping the $B$ and $B'$, gives the input to the decoder as  
\begin{equation}
\label{ew:swap_lat}
\ket{\chi'}=\mathbf{I}_A\otimes\mathcal{V}_{BB'}\;\; \ket{\chi}_{AB}\otimes \ket{\beta}_{B'}~,
\end{equation}where $\mathcal{V}_{BB'}$ indicates a unitary that performs the swap operation,\footnote{For instance swapping the state of two qubits  in the basis $\{\ket{00},\ket{01},\ket{10},\ket{11}\}$, can be implemented via the unitary matrix
$$\mathcal{V}_{BB'}=\begin{bmatrix} 
1&0&0&0\\ 
0&0&1&0\\
0&1&0&0\\
0&0&0&1
\end{bmatrix} \quad.$$
} and $\mathbf{I}_A$ is the identity operator on $\mathcal{H}_A$. The output of the decoder can now be written as 
\begin{equation}
\label{eq:dec_quant}
\ket{\Psi}=\mathbf{U}^\dagger_{AB}(\Theta)\otimes\mathbf{I}_{B'} \;\ket{\chi'}~,
\end{equation}with $\mathbf{I}_{B'}$ being the identity operator on $\mathcal{H}_{B'}$. The decoding, therefore, takes the swapped latent state $\ket{\chi'}$, and the unitary $\mathbf{U}^\dagger_{AB}$ evolves it with no information from the encoder in the subspace $\mathcal{H}_B$. The reconstruction efficiency of the autoencoder can be quantified in terms of the \emph{fidelity} between the input and output states in the subspace $\mathcal{H}_A\otimes\mathcal{H}_B$, which quantifies their similarity. For two quantum states $\ket{\psi}$ and $\ket{\phi}$, it is defined as $$F(|\phi\rangle,|\psi\rangle)=F(\ket{\psi},\ket{\phi})=|\braket{\phi}{\psi}|^2~.$$ For normalized states, we have $0\leq F\leq1$, with $F=1$ only when $\ket{\phi}$ and $\ket{\psi}$ are exactly identical. We can write the fidelity of the complete system as 
\begin{equation*}
\begin{split} 
F&(\ket{\Phi}_{AB}\otimes\ket{\beta}_{B'},\ket{\Psi})\\&=F(\ket{\Phi}_{AB}\otimes\ket{\beta}_{B'},\mathbf{U}^\dagger_{AB}\;\mathcal{V}_{BB'} \;\mathbf{U}_{AB}\;\ket{\Phi}_{AB}\otimes\ket{\beta}_{B'})~,
\end{split} 
\end{equation*}where we have implicitly assumed that the unitary operators are extended to the whole space via a direct product with the identity operator on the subspace it does not act on, for notational compactness. Noting that $\mathbf{U}_{AB}\ket{\Phi}_{AB}=\ket{\chi}_{AB}$, we can write this as,
\begin{equation*} \begin{split} 
F&(\ket{\Phi}_{AB}\otimes\ket{\beta}_{B'},\ket{\Psi})\\
&=F(\ket{\chi}_{AB}\otimes\ket{\beta}_{B'},\mathcal{V}_{BB'} \;\ket{\chi}_{AB}\otimes\ket{\beta}_{B'})~.
\end{split} 
\end{equation*}  Writing the swapped state as $\mathcal{V}_{BB'}\;\ket{\chi}_{AB}\otimes\ket{\beta}_{B'}=\ket{\chi}_{AB'}\otimes\ket{\beta}_{B}$, we have 
\begin{equation} \label{eq:fid}
F(\ket{\Phi}_{AB}\otimes\ket{\beta}_{B'},\ket{\Psi})=F(\ket{\chi}_{AB}\otimes\ket{\beta}_{B'},\ket{\chi}_{AB'}\otimes\ket{\beta}_{B})~.
\end{equation}Since we are interested in the wave functions belonging to the subspace $\mathcal{H}_A\otimes\mathcal{H}_B$, we trace over $B'$ to get the required fidelity. However, a perfect fidelity between the input and outputs of the $AB$ system can be achieved when the complete information of the input state passes to the decoder, i.e. 
\begin{equation}
\label{eq:trash_decomposition}
\mathbf{U}_{AB}\ket{\Phi}_{AB}=\ket{\Phi^c}_A\otimes\ket{\beta}_B\quad.
\end{equation} The state $\ket{\Phi^c}_A$ denotes a compressed form of $\ket{\Phi}_{AB}$, i.e it should contain the information of the $AB$ system in the input, while $\ket{\beta}_{B}$ is equivalent to the reference state, with no information of the input. If the $B$ and $B'$ systems are identical during the swap operation, the entire circuit reduces to the identity map.  The output of the $B'$ system, hereby referred to as the \emph{trash state}, is itself the determining factor of the output state fidelity. The output of the $B'$ system can be obtained after tracing over the $A$ system as: $\hat{\rho}_{B'}=\trace{A}{\ket{\chi}\bra{\chi}_{AB'}}$ and the required fidelity of the $B'$ system is $F(\ket{\beta}_{B'},\hat{\rho}_B')$.

A perfect reconstruction of the input is possible only when the trash state fidelity $F(\ket{\beta}_{B'},\hat{\rho}_B')=1$. Thus a quantum autoencoder can be trained by maximising the trash state fidelity instead of the output fidelity, which has the advantage of reducing the resource requirements during training. Although, the output fidelity obtained by tracing over the $B'$ system is numerically not equal to the trash state fidelity, we can use the latter in anomaly detection as well, since it is a faithful measurement of the output fidelity. Thus, unlike vanilla classical autoencoders, we can reduce the execution and training of QAEs into the encoder circuit for anomaly detection.  

The above discussions have focused on the underlying principles behind a quantum autoencoding process on single input states. As stated before, we need to prepare identical input-states for each data point and repeat the unitary evolution and measurement to get a useful estimate of the fidelity, evident also from the use of density operators to express the output state. Referring to the ensemble of the input states as $\{p_i,\ket{\Phi_i}_{AB}\}$, we obtain for the cost function  
\begin{equation}
\label{eq:qae_cost}
C(\Theta)=-\sum_{i} p_i\;F(\ket{\beta}_{B'},\hat{\rho}_B')\quad,
\end{equation}
where the negative sign converts the optimisation process into minimising the cost function. It is important to note that the ensemble should not be taken as being analogous to the batch training in classical neural networks, as it is required for the accurate prediction of the network output even when testing the autoencoder network. 
  
\section{Analysis Setup}
\label{sec:an_setup} 
\begin{figure}
	\centering
	\includegraphics[scale=0.25]{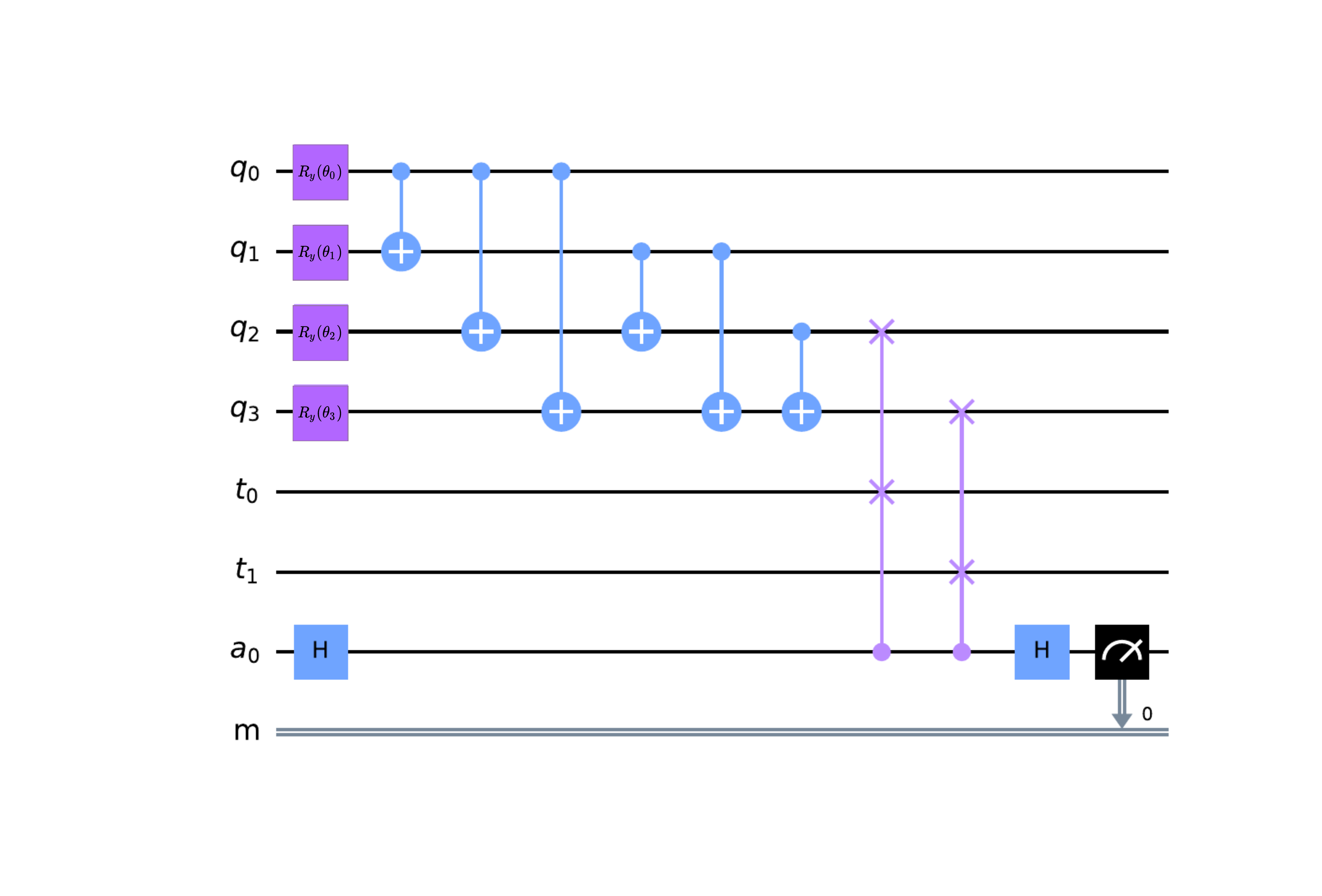}
	
	\caption{The figure shows a Quantum autoencoder circuit for a four qubit input and two latent qubits. The inputs are already embedded in $q_i$ (by the input embedding circuit),  which are then rotated by tunable angles $\theta_i$ in the y-direction of the Bloch sphere by $R_y(\theta_i)$ gates (shown in purple boxes). Each pair of these qubits are entangled via CNOT gates(shown with blue lines). For the trash training, we need a two-dimensional reference state denoted by $t_i$ qubits and an ancillary qubit $\texttt{a}_0$. The fidelity between two qubits at the encoder output and the reference states is measured via a SWAP test.}
	\label{fig:q_circuit}
\end{figure}
\subsection{Data simulation}
\label{sec:data_sim}

	To show the prowess of the quantum autoencoders, we study two processes with distinctive features: a QCD continuum background of top pair production taking possible signal signatures of resonant heavy Higgs decaying to a pair of top quarks, and invisible Z decays into neutrinos with a likely signal of the 125 GeV Higgs decaying to two dark matter particles. As we shall see in the following sections, the relative performance of QAEs over CAEs show parallels in these two different signatures, pointing towards an advantage of QAEs over CAEs not governed by the specific details of the final state. 
	
\subsubsection{Resonant Higgs signal over continuum $t\bar{t}$ background}
The first background and signal samples used in our analysis consist of the QCD $t\bar{t}$ continuum production, 
$pp \to t\bar{t}$, and the scalar resonance production $pp \to H \to t\bar{t}$, respectively.
The background and the signal events are generated with a centre-of-mass energy of 14~TeV, as expected during future LHC runs.
Each top decays to a bottom quark and a $W$ boson, and we focus on the decay of the $W$'s into muons exclusively. We consider four different masses of the scalar resonance, $m_{H} = 1.0, 1.5, 2.0$, and $2.5$~TeV. All events are generated with {\tt MadGraph5\_aMC@NLO}~\cite{Alwall:2014hca}, and showered and hadronization is performed by {\tt Pythia8}~\cite{Sjostrand:2014zea}. 
{\tt Delphes3}~\cite{deFavereau:2013fsa} is utilized for the detector simulation, where the jets are clustered using {\tt FastJet}~\cite{Cacciari:2011ma}. We generate 
about 30k events for the background samples, while for each signal sample, we generate about 15k.
The background events are divided into 10k training, 5k validation and 15k testing samples. 

For the object reconstruction, a standard jet definition using the anti-$k_t$ algorithm~\cite{Cacciari:2008gp} with the jet radius $R=0.5$ is used. For the signal bottom jets, the output from Delphes 3 is used and require $p_{T}^b > 30$~GeV. 
For isolated leptons, we requires $p_{T}^l > 30$~GeV and its isolation criteria with $R=0.5$. We extracted four variables $\{p^{b_1}_T,p^{l_1}_T,p^{l_2}_T,\slashed{E}_T\}$ for our analysis, keeping in mind the limitations of current devices. To conserve the aperiodic topology of these variables in the angle embedding (given in eq.~\ref{eq:enc_quantum}) we fix the range of each variable to $[0,1000]$ by adding two points\footnote{Events with the variables lying above 1000 GeV are very rare and excluded in our case. In a realistic analysis, the upper bound can be determined from the data.} and map the whole dataset to a range $[0,\pi]$ via the {\tt MinMaxScaler} implemented in {\tt scikit-learn}~\cite{scikit-learn}. The two added points are then removed from the dataset. This maps each feature's minimum and maximum to two distinct angles separated by a finite distance due to the selection criteria. 

\subsubsection{Invisible Higgs signal over invisible $Z$ background}
To test the anomaly detection capabilities of QAEs in a different scenario, we study invisible decays of a Z boson produced with two jets originating from QCD vertices. As a possible signal, we take the production of the 125 GeV Higgs boson and two jets originating from Electroweak vertices, decaying to two scalar dark matter particles. The generation is carried out in the same manner as in the previous case, including the definition of jets. We demand that we have at least two reconstructed jets with $p_T>30 $ GeV, and the events have a missing transverse momentum $\slashed{E}_T>30$ GeV. For the background, we have 30k events divided into 10k training, 5k validation, and 15k test events, while for the signal, we have 15k test events. We extract six variables to train the QAE and the CAE. They are the absolute separation in pseudorapidity between the two jets $|\Delta \eta_{jj}|$, the invariant mass of the dijet system $m_{jj}$ and the sum of transverse energies $$H^{\eta_C}_T=\sum_{|\eta_i|<\eta_C}\; E^i_T\quad,$$ within four ranges of pseudorapidity $\eta_C\in\{1.0,1.5,2.0,2.5\}$. The mapping to conserve the aperiodic topology of these variables in the angular embedding is done by increasing their range on the higher side.

\subsection{Network architecture and training}
The QAE was implemented and trained using {\tt Pennylane}~\cite{bergholm2018pennylane}. As stated before, we train and test the QAE model with only the encoder circuit. After the input features are embedded as the rotation angle of the x-axis in the Bloch sphere, the unitary evolution $\mathbf{U}(\Theta)$ consists of two stages. In the first step, each qubit is rotated by an angle $\theta_i$ in the y-axis of the Bloch sphere. The values of these angles are to be optimized via gradient descent. After this, we apply the CNOT gate to all the possible pairs of qubits, with the ordering determined by the explicit number of the qubit. This circuit is shown in figure~\ref{fig:q_circuit} for a four qubit input QAE with two-qubit latent representation. It is given by,
\begin{equation*}
\mathbf{U}_{AB}= C_{23}\otimes C_{13}\otimes C_{12}\otimes C_{23}\otimes C_{03}\otimes C_{02}\otimes C_{01}\otimes R_y^0(\theta_0)\otimes R_y^1(\theta_1)\otimes R_y^2(\theta_2)\otimes R_y^3(\theta_3)\quad,
\end{equation*} 
where $C_{ij}$ is the CNOT operation acting on the composite space of two qubits $i$ and $j$, and $R_y^i(\theta_i)$ is the rotation of a single qubit $i$ about the y-axis of the Bloch sphere. Note that the expression does not contain the operations of the SWAP test, which will be explained in the following paragraphs. The training proceeds to find the optimal values for $\theta_i$.

The number of qubits discarded at the encoder, the size of the trash-state, fixes the latent dimension\footnote{In our discussion, we will use the number of latent qubits as the latent dimension, although the Hilbert space would have $2^{N_{lat}}$ dimensions.} via $N_{lat}=N_{in}-N_{trash}$, with $N_{lat}$ the latent dimension, $N_{in}$ the size of the input state, and $N_{trash}$ the number of discarded qubits. The reference state $\ket{\beta}_{B'}$, has the same number of qubits $N_{trash}$, and it is initialized to be $$\ket{\beta}_{B'}=\ket{0}^{\otimes N_{trash}}\quad.$$ We measure the fidelity between the trash-state $\hat{\rho}_{B'}$ and the reference state $\ket{\beta}_{B'}$ via a SWAP test~\cite{PhysRevLett.87.167902}. It is a way to measure the fidelity between two multi-qubit states. For any two states $\ket{\phi}$ and $\ket{\psi}$ with the same dimensions, the fidelity $F(\ket{\phi},\ket{\psi})$ can be measured as the output of an ancillary qubit $\ket{a}_{anc}$ after the following operation,
\begin{equation}
\label{eq:fid_anc}
\mathbf{H}_{anc}\otimes \mathbf{I}\;(\text{c-SWAP})\;\mathbf{H}_{anc}\otimes \mathbf{I} \;\;\ket{0}_{anc}\otimes\ket{\phi}\otimes\ket{\psi}\quad,
\end{equation}where $\mathbf{H}_{anc}$ is the Hadamard gate acting on the ancillary qubit, and c-SWAP is the controlled swap operation between the states $\ket{\phi}$ and $\ket{\psi}$ controlled by the ancillary qubit. Thus the total number of qubits required for a fixed $N_{in}$ and $N_{trash}$ is $N_{in}+N_{trash}+1$. Due to the limitation of current quantum devices we limit the input feature to four, and scan over the possible latent dimensions. 

\begin{figure*}[th!]
	\centering
	\includegraphics[scale=0.19]{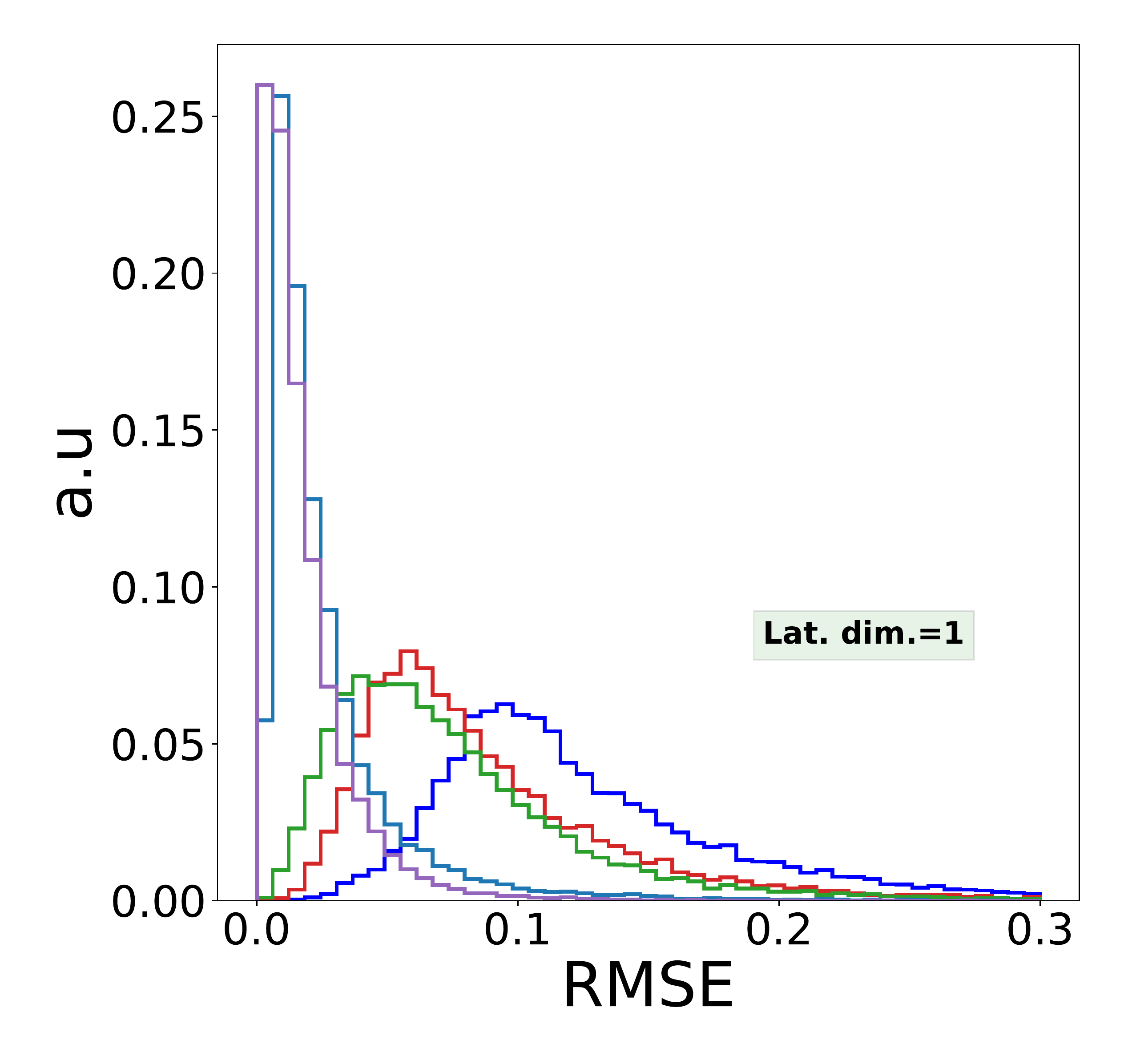}
	\includegraphics[scale=0.19]{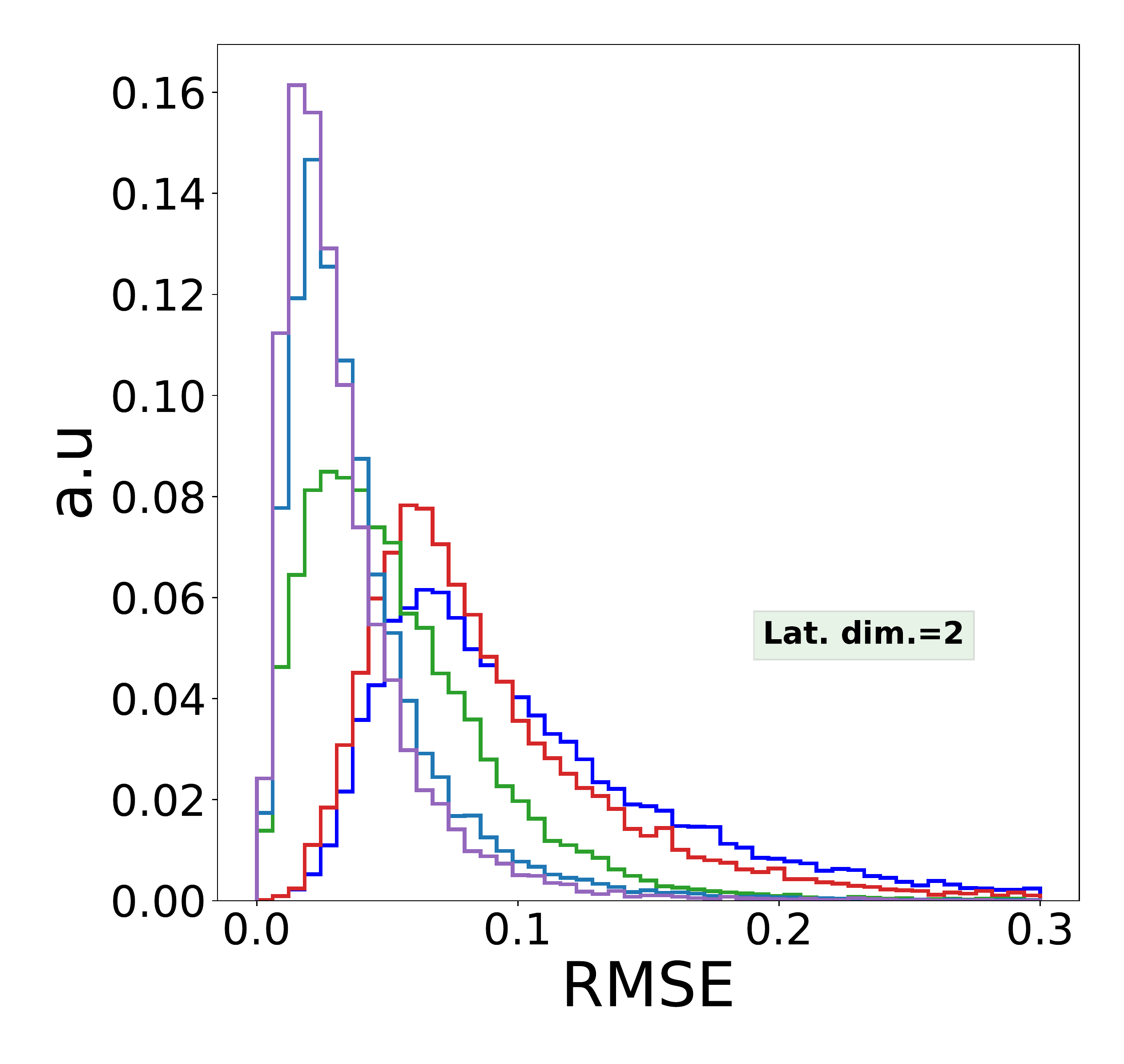}
	\includegraphics[scale=0.19]{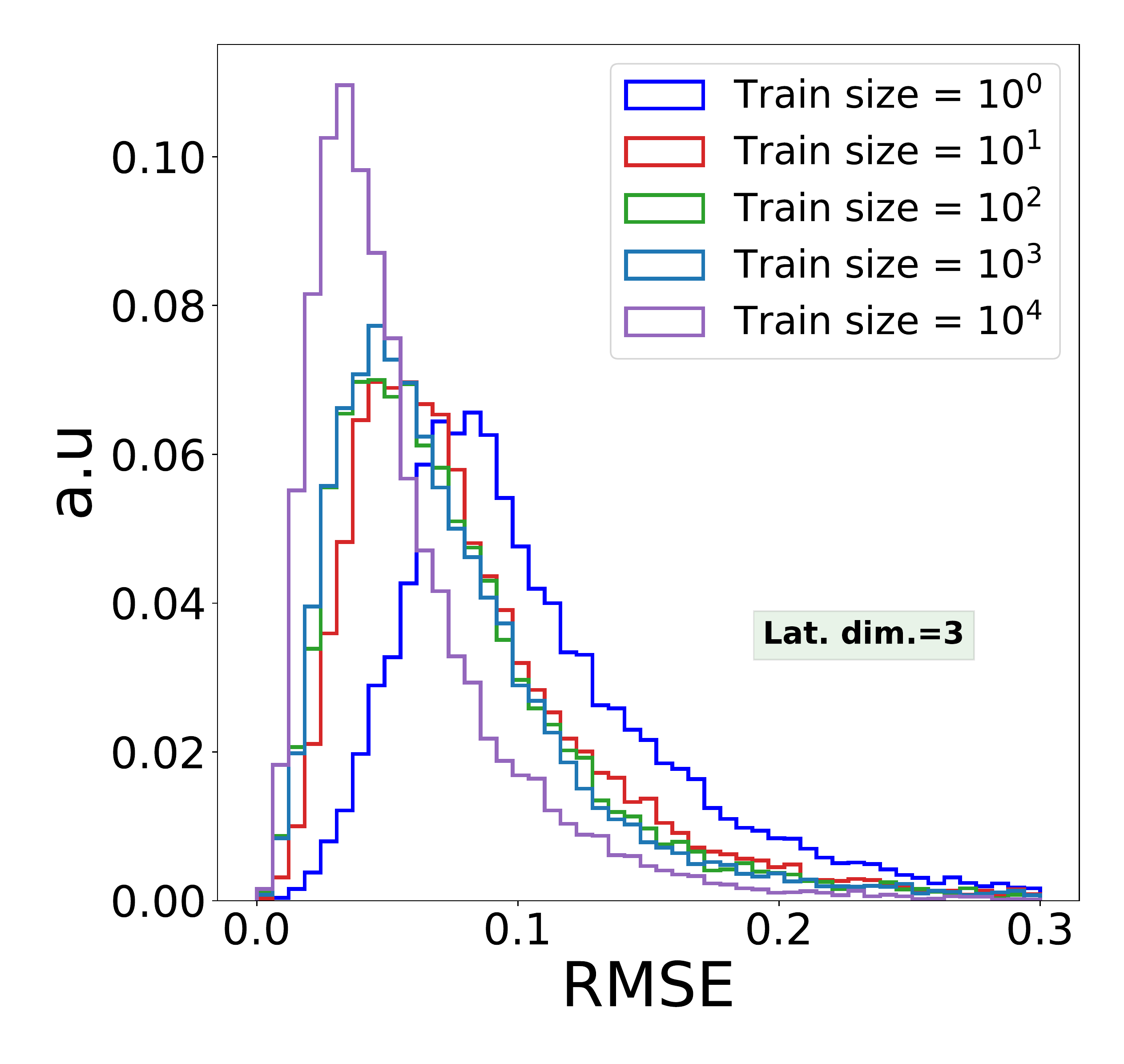}	
	\caption{Loss distribution of the test background samples (15k) for different sizes of training dataset. We can see that the distribution shifts significantly towards the left (direction with lower loss) as one increases the training data size, which reflects that there is noticeable increase in learning with larger data samples.}
	\label{fig:hist_cae} 
\end{figure*}  
\begin{figure*}[th]
	\centering
	\includegraphics[scale=0.19]{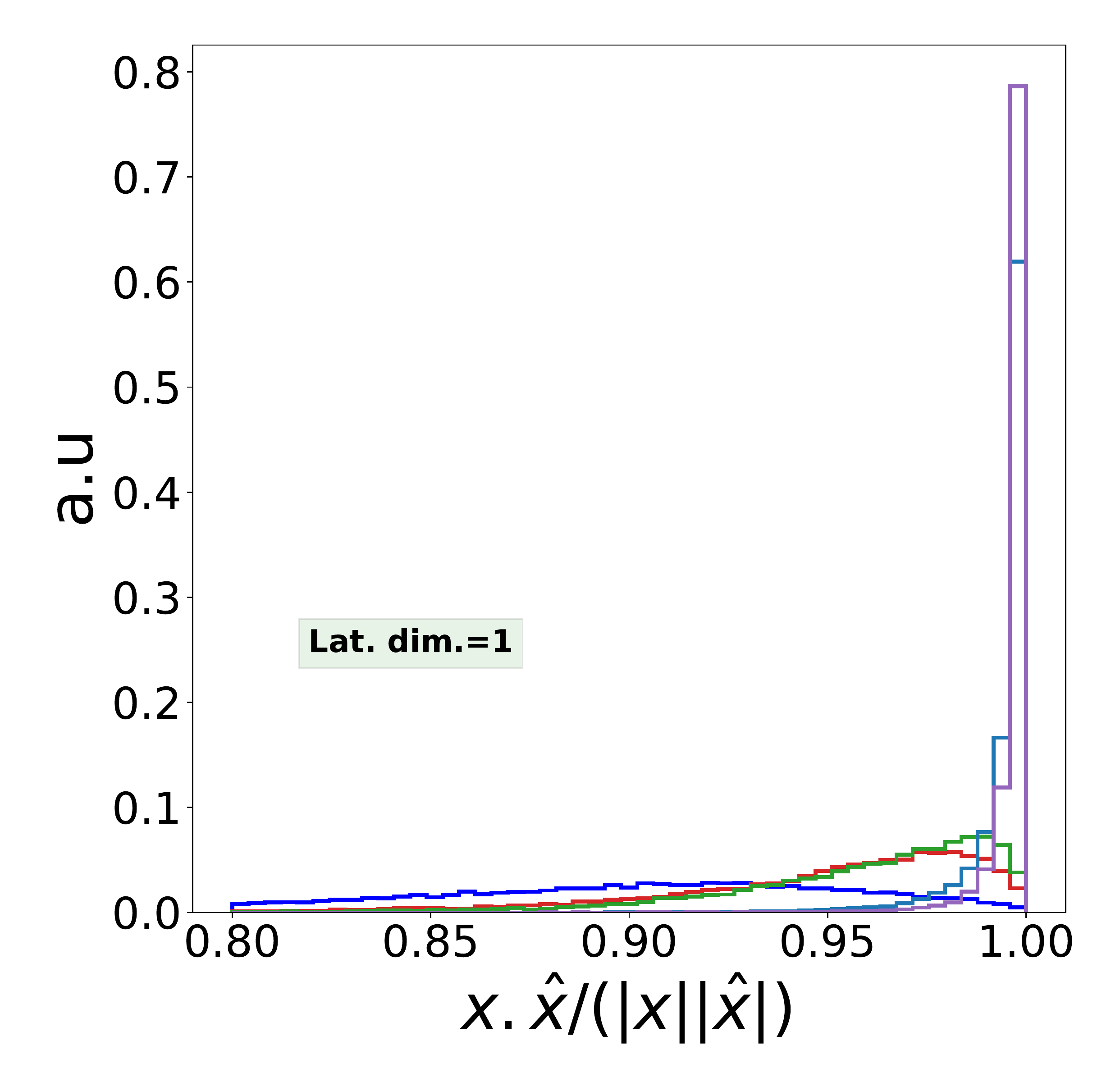}
	\includegraphics[scale=0.19]{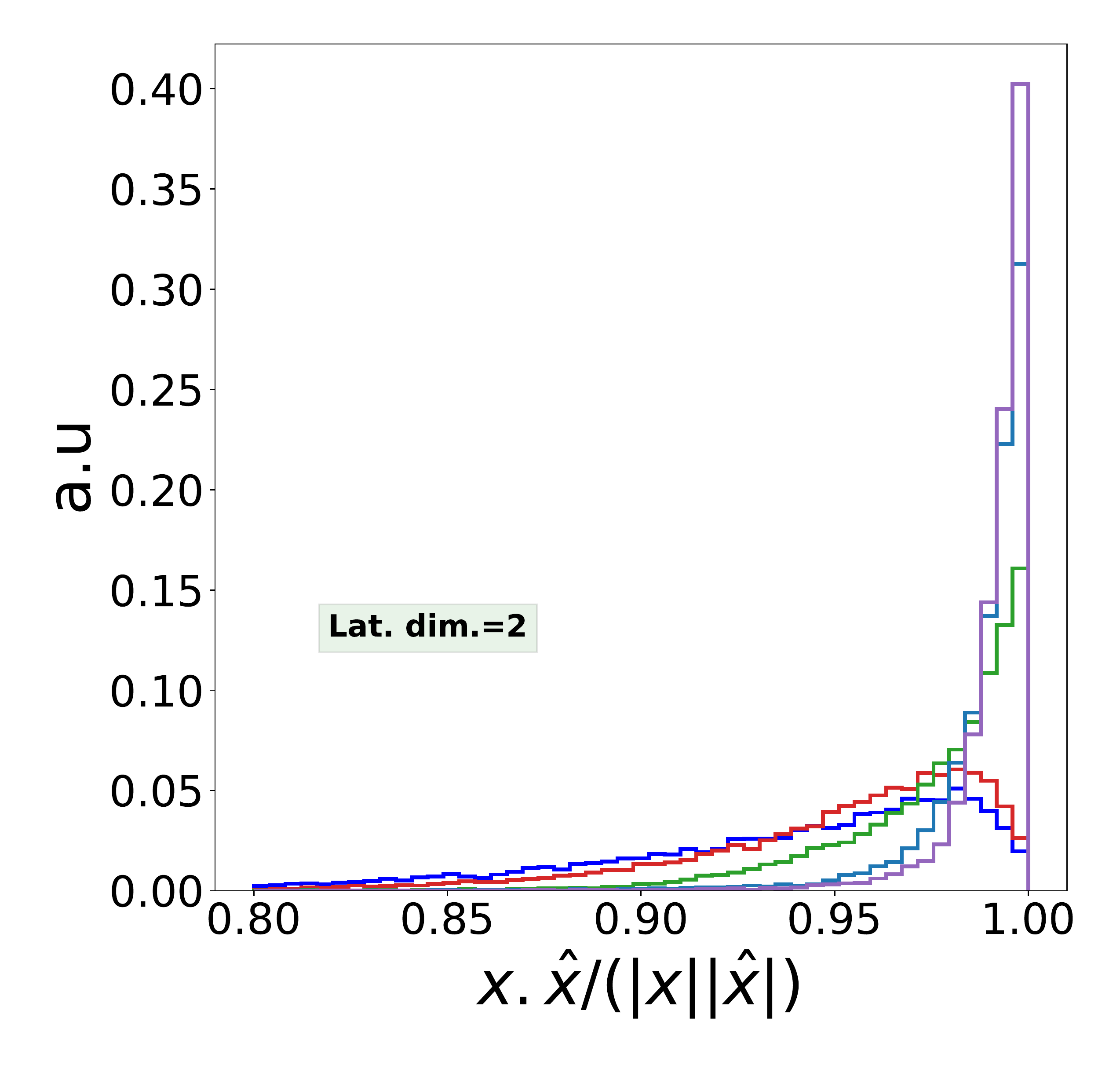}
	\includegraphics[scale=0.19]{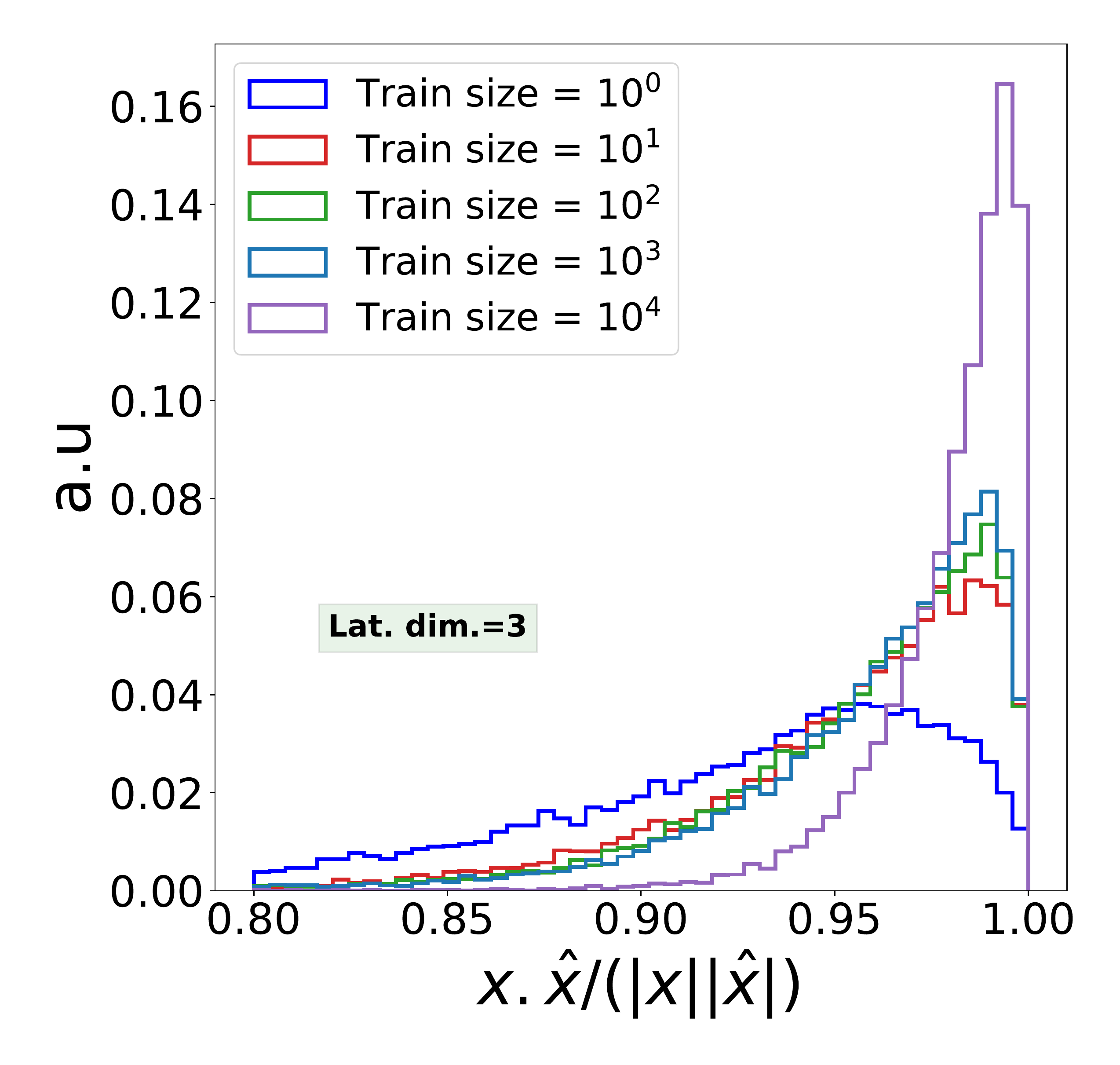}	
	\caption{Cosine similarity (analogous to quantum fidelity) distribution of the test background samples (15k) for different sizes of training dataset of the CAE. }
	\label{fig:hist_cae_fid} 
\end{figure*}
\begin{figure*}[th!]
	\centering
	\includegraphics[scale=0.19]{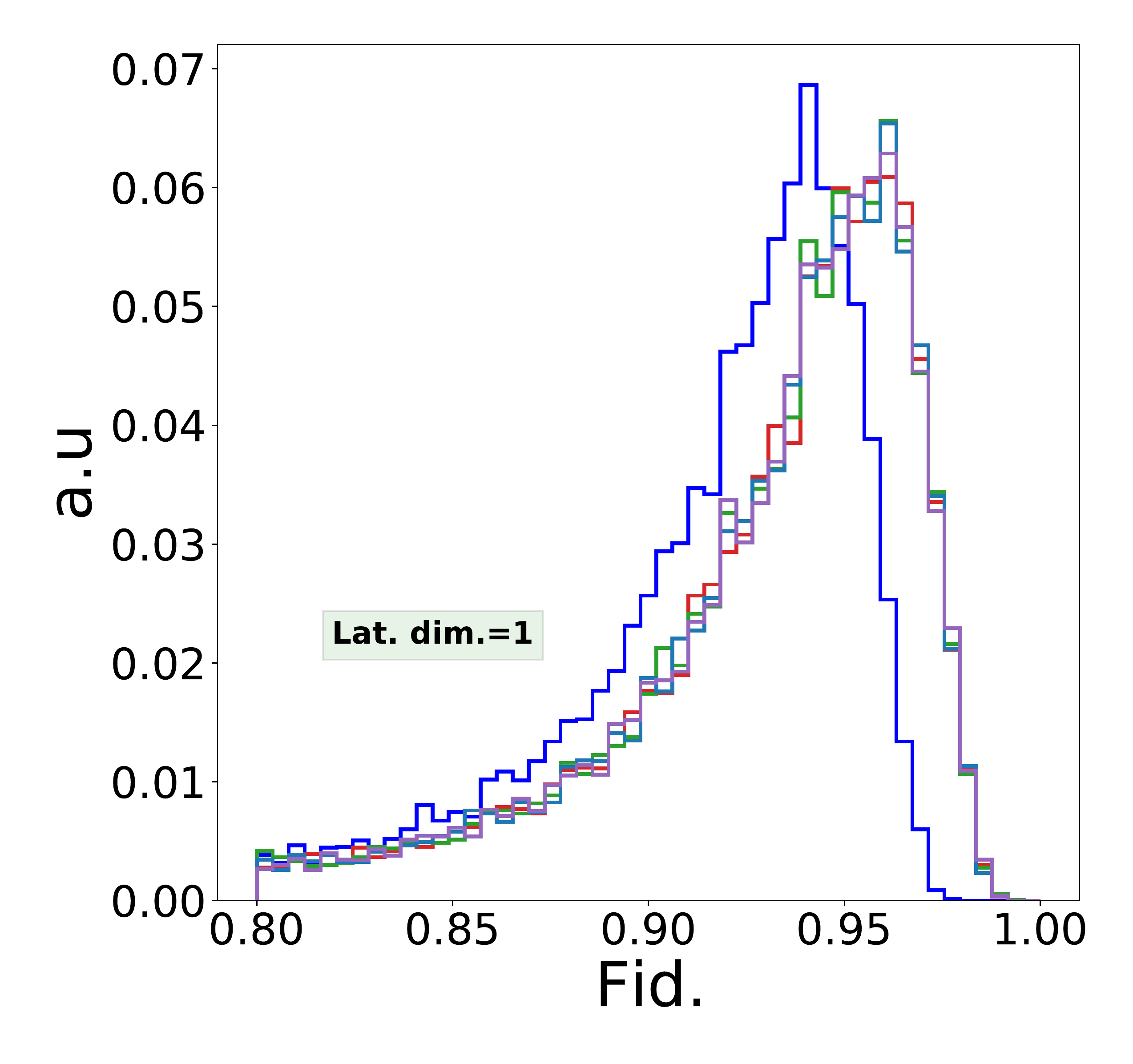}
	\includegraphics[scale=0.19]{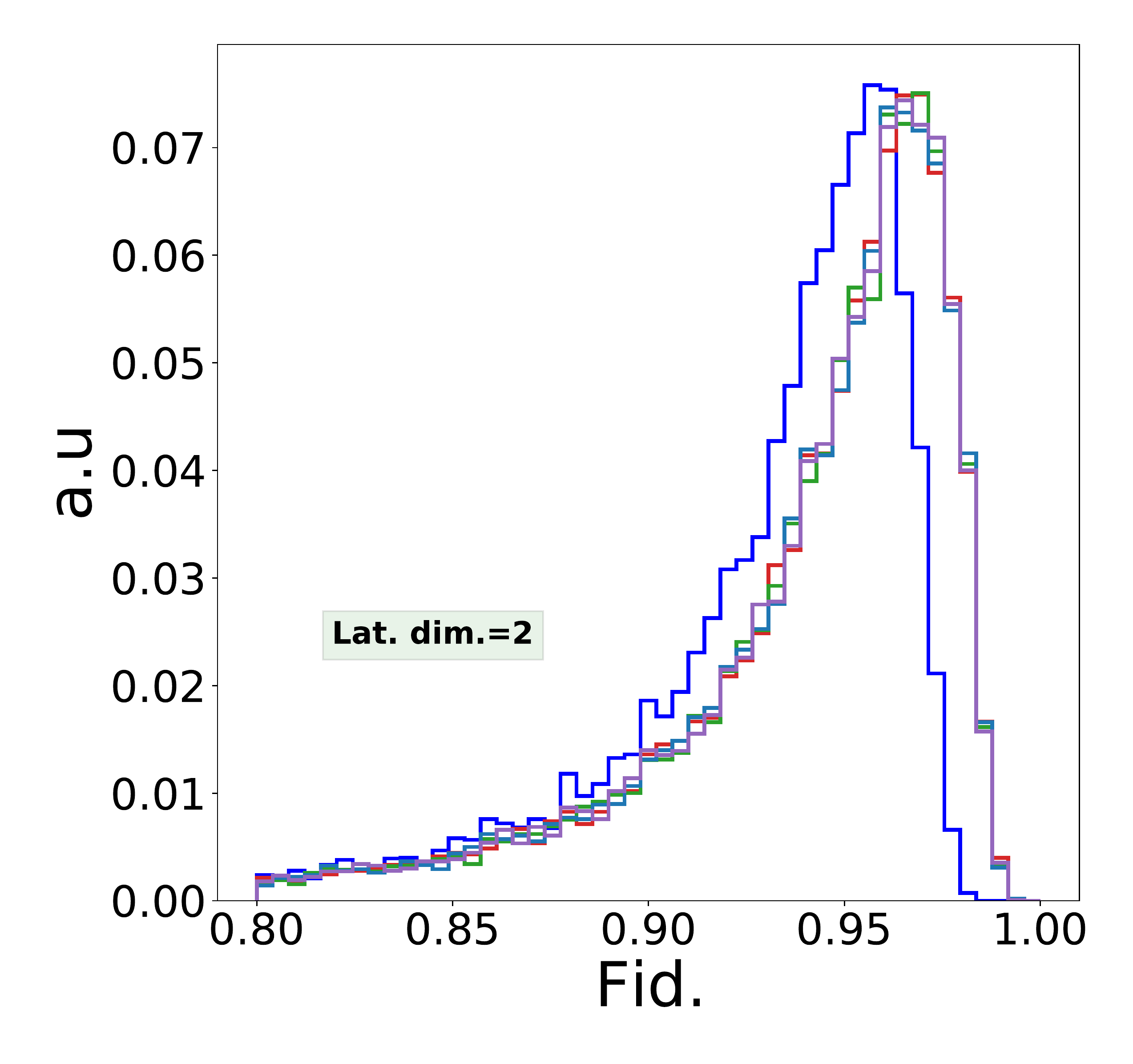}
	\includegraphics[scale=0.19]{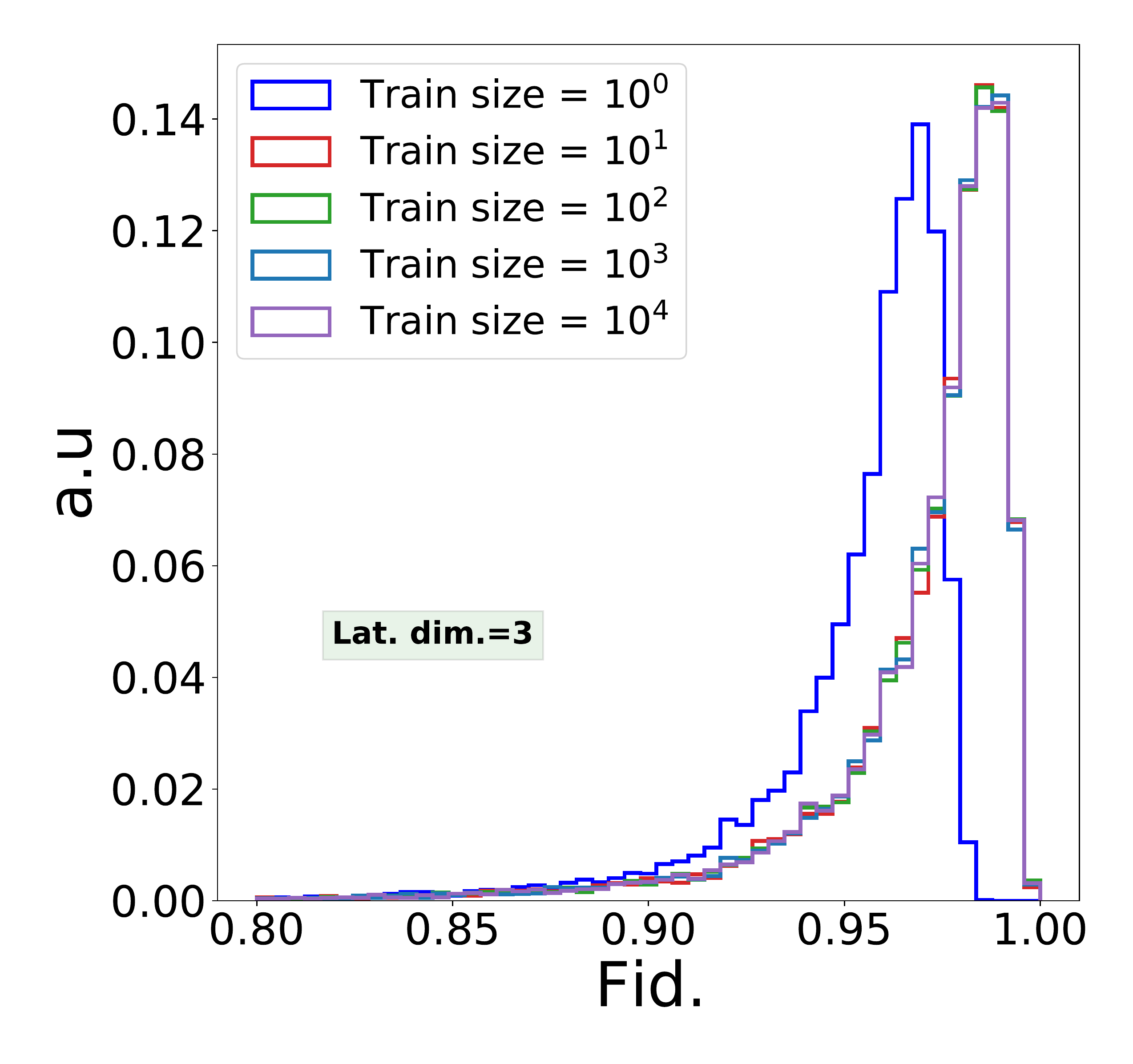}	
	\caption{Quantum Fidelity distribution of the test background samples (15k) for different sizes of training dataset for the QAE. The peak shifts towards the right in analogy to the CAE, however the shift is not as pronounced. With a single training sample, the network is not able to converge completely while for anything greater than 10, the increase in training size has practically no effect. }
	\label{fig:hist_qae} 
\end{figure*}

The quantum network is trained by minimising the cost function (c.f eq.~\ref{eq:qae_cost}) with quantum gradient descent for the one, two and three-dimensional latent spaces. We train these instances for different training sizes of 1, 10, 100, 1000, and 10000 events to study the dependence of the QAE's performance on the size of the training data. We update the weights for each data sample, with 5000 shots in all training scenarios. For training sizes greater than or equal to 100, we train the networks for 50 epochs. In comparison, for sample sizes 1 and 10, we train the QAE for 500 and 200 epochs, respectively.
To benchmark the performance of a QAE on a quantum computer, we train a QAE with the two inputs $p_T^{l_1}$ and $p_T^{b_1}$ with quantum-gradient descent on {\tt Pennylane}, and compare the test performance with the simulation and the IBM-Q. For running on the IBM-Q, we build and implement the test circuits in {\tt Qiskit}~\cite{Qiskit}.

We also train classical autoencoders using {\tt Keras-v2.4.0}~\cite{chollet2015keras} with {\tt Tensorflow-v2.4.1}~\cite{tensorflow2015-whitepaper} for the same input features, for comparison. The encoder is a dense network mapping the input space to a latent dimension of $N_{lat}\in\{1,2,3\}$, and has three hidden layers with 20, 15, and 10 nodes. 
%\vn{Keeping the number of layers and nodes fixed, we run a {\tt RandomSearch} hyperparameter scan using {\tt KerasTuner-v1.1.0}\footnote{The details are elucidated in appendix \ref{app:hyper_tune} and choose the other parameters accordingly.}}
The hidden layers have {\tt ReLU} activations while the latent output has {\tt Linear} activation. The decoder has a symmetric configuration to the encoder. The networks are trained with {\tt Adam}~\cite{kingma2017adam} optimiser with a learning-rate of $10^{-3}$ to minimise the root-mean-squared error between the input vector $\mathbf{x}$ and the reconstructed vector $\hat{\mathbf{x}}$. For the CAEs, we found that training with single data per update (technically batch size=1) has a volatile validation loss per epoch, with slow convergence. Therefore, we choose a batch size of 64 to train the CAEs.\footnote{The network performs one update per epoch for training with a number of samples less than 64. These training sizes are too small for a CAE to have any good learning capability. Hence, we do not try to modify the batch sizes or interpret the test distributions.}

We train the QAE with analogous architecture for a six-dimensional input for the second scenario for a two-dimensional latent space in a similar fashion for all training sizes. For the CAE keeping the number of nodes and layers identical to the previous case for six-dimensional input and output vectors, we perform a hyperparameter scan, the details of which is given in Appendix~\ref{app:hyp_scan}. All results shown in the next section for this scenario is for the best performing hyperparameters.

\section{Results}
\label{sec:results}

Results of the various training scenarios are presented in this section. We present a detailed investigation of the QAE and CAE's properties for the $t\bar{t}$ background scenario in Sections~\ref{sec:tt_train_dep} to \ref{sec:tt_ibmq}. The lessons learnt from these analyses, particularly the training size independence and the relative performance, are then tested for the invisible $Z$ background in Section~\ref{sec:z_inv_comp}.

\subsection{Dependence of test reconstruction efficiency on the number of training samples}
\label{sec:tt_train_dep}
The distribution of the loss function of the independent background test samples for different training sizes of the CAE is shown in figure~\ref{fig:hist_cae}. Although training with a single data point is inherently inaccurate, we perform such an exercise as a sanity check of the CAE's comparison to a QAE. The test distribution shifts towards the left as one increases the training size, thereby signifying increased reconstruction efficiency.  For training sizes of up to $10^2$, the limited statistics will produce a very high statistical uncertainty. Since it is not the main emphasis of our present work, we do not comment any further.  Looking at the distribution across different latent dimensions for $10^3$ and $10^4$ training samples, one can see the impact of the information bottleneck. For a singular latent dimension, the passed information is already available from $10^3$ samples, and hence the loss distribution is very close to the one trained on $10^4$. This relative separation increases as we go to higher latent dimensions, denoting the higher information passed to the decoder to reconstruct the input, which is exploited with higher training samples. For an analogous comparison with the quantum fidelity, we define the cosine similarity between the input vector $\mathbf{x}$ and the reconstructed vector $\hat{\mathbf{x}}$  as,
\begin{equation}
\label{eq:cos_sim}
\cos\alpha=\frac{\mathbf{x}.\hat{\mathbf{x}}}{|\mathbf{x}|\;|\hat{\mathbf{x}}|}~,
\end{equation} where the dot product is done with a Euclidean signature. The distribution of the cosine similarity shown in figure~\ref{fig:hist_cae_fid}, shows similar features to the loss function's distribution, with efficient reconstruction possible only when the train size is at least $10^3$. 

\begin{figure*}[t]
	\centering
	\includegraphics[scale=0.25]{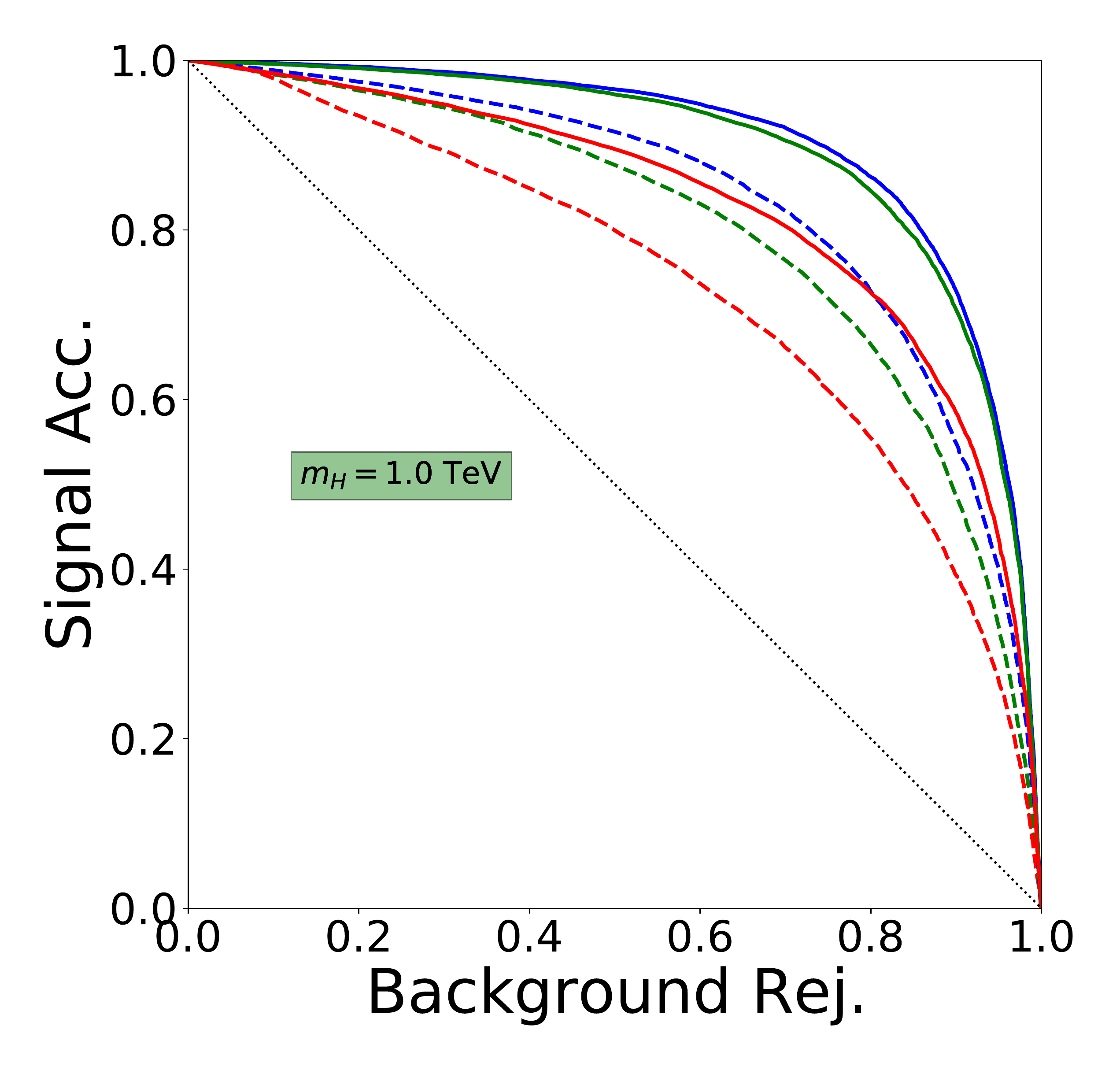}
	\includegraphics[scale=0.25]{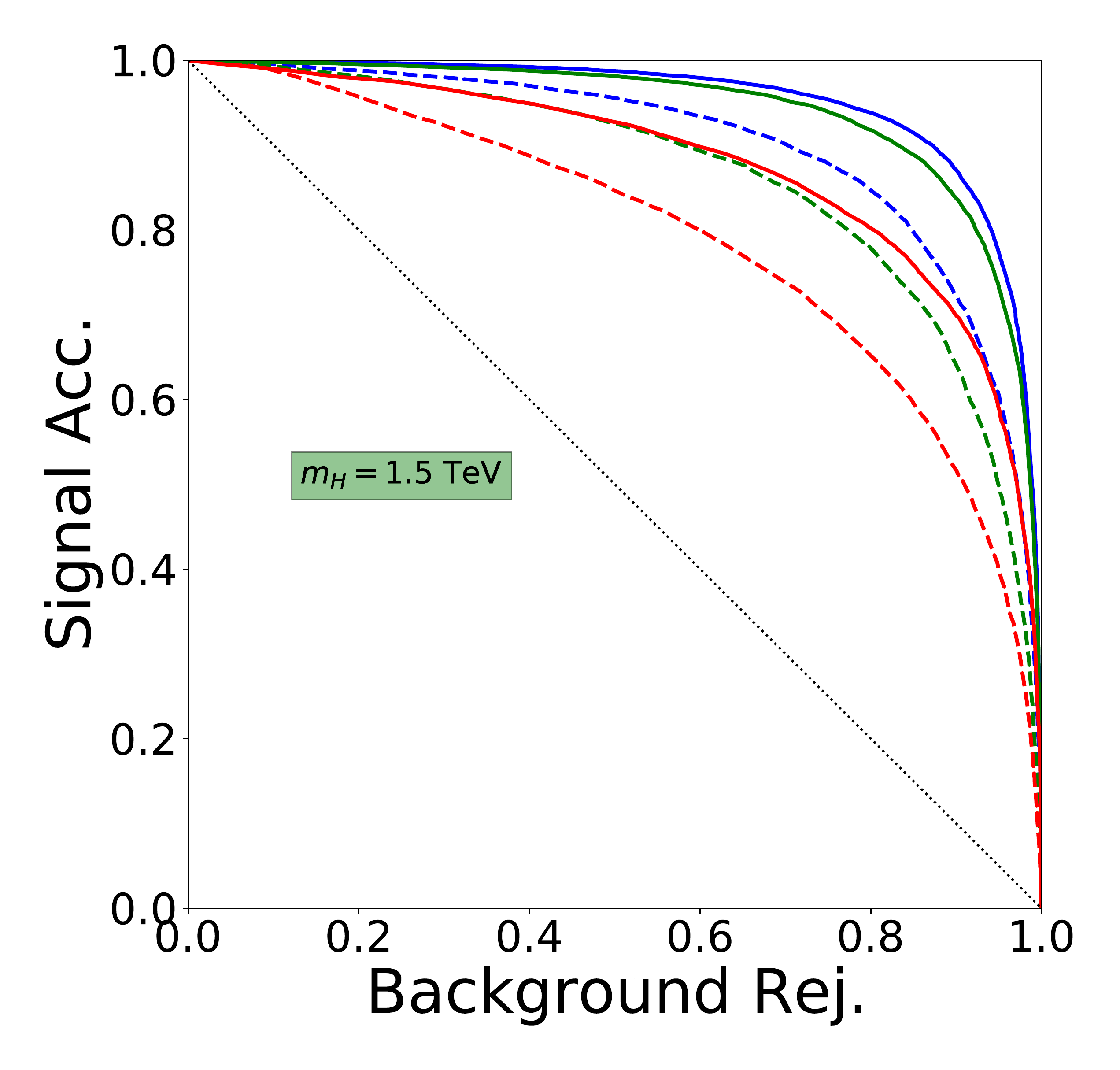}\\
	\includegraphics[scale=0.25]{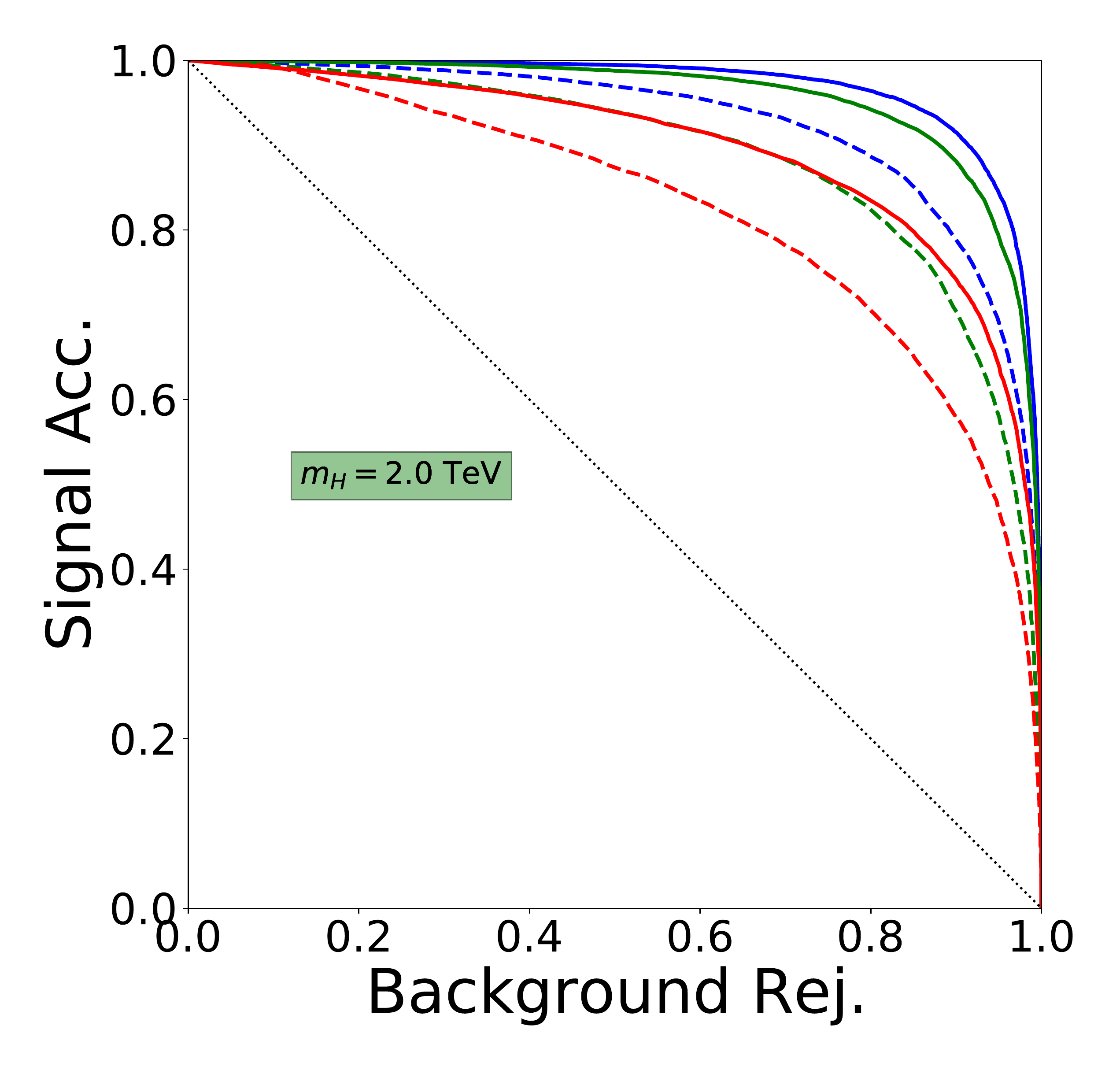}
	\includegraphics[scale=0.25]{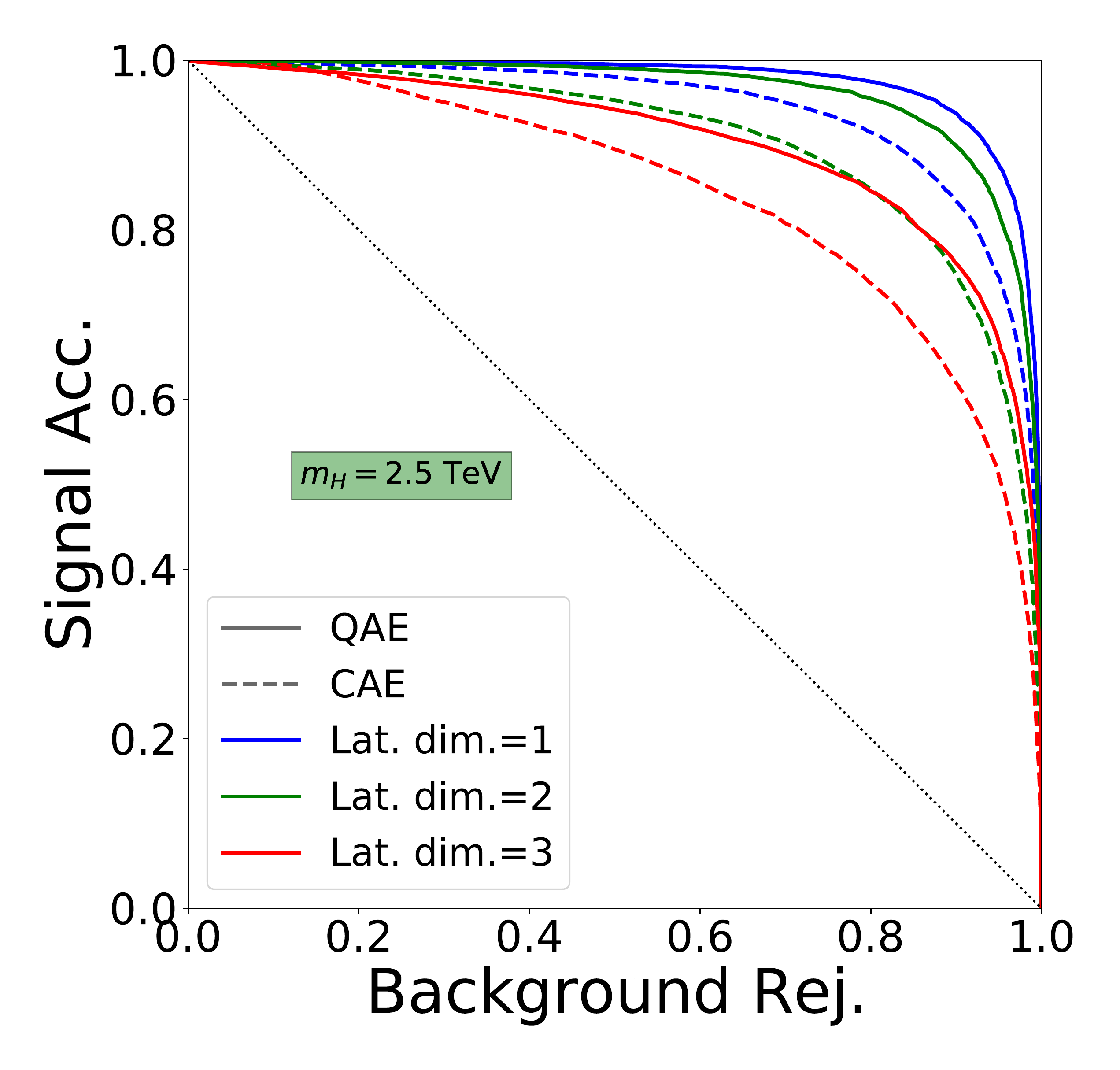}	
	\caption{ROC curve between signal acceptance vs background rejection for Quantum Autoencoder (QAE) and Classical Autoencoder (CAE) for various values of $m_H$  and different latent dimensions for a training datasize of 10k samples. The trend across latent dimensions is same for both QAE and CAE with QAEs performing better in all cases.}
	\label{fig:roc} 
\end{figure*}

We have seen that CAEs cannot be trained with limited statistics to reconstruct the statistically independent test dataset. From the distribution of the test sample's fidelity in figure~\ref{fig:hist_qae}, we see that QAEs are much more effective in learning from small data samples. Although training with a single data point has not reached the optimal reconstruction efficiency,  it is obtained with ten sample events. Unlike CAEs, see figs.~\ref{fig:hist_cae} and \ref{fig:hist_cae_fid}, the test fidelity distribution for all latent spaces are identical for training sizes greater than or equal to ten. The independence of the sample size is particularly important in LHC searches where the background cross section is small. This particularly
	interesting feature may be due to the interplay of an enhancement of statistics via the uncertainty of quantum
	measurements and the relatively simple circuits employed in our QAE circuit. For a single input point and assuming
	that we have hardware capable of building exact copies, a finite number of measurement processes always introduces
	a non-zero uncertainty in the network output. This uncertainty can act as additional information in the quantum
	gradient minimisation, which is performed after the measurement process, increasing the convergence for smaller data
	samples. Moreover, existing studies~\cite{qml_adv,Huang2021} show the advantage of quantum machine learning over classical approaches. Additionally, the use of quantum gradient descent~\cite{Blance2021} makes the loss landscape more convex, thereby speeding up convergence.

\subsection{Classification Performance}
\label{sec:tt_class_perf}
\begin{figure*}[t]
	\centering
	\includegraphics[scale=0.2]{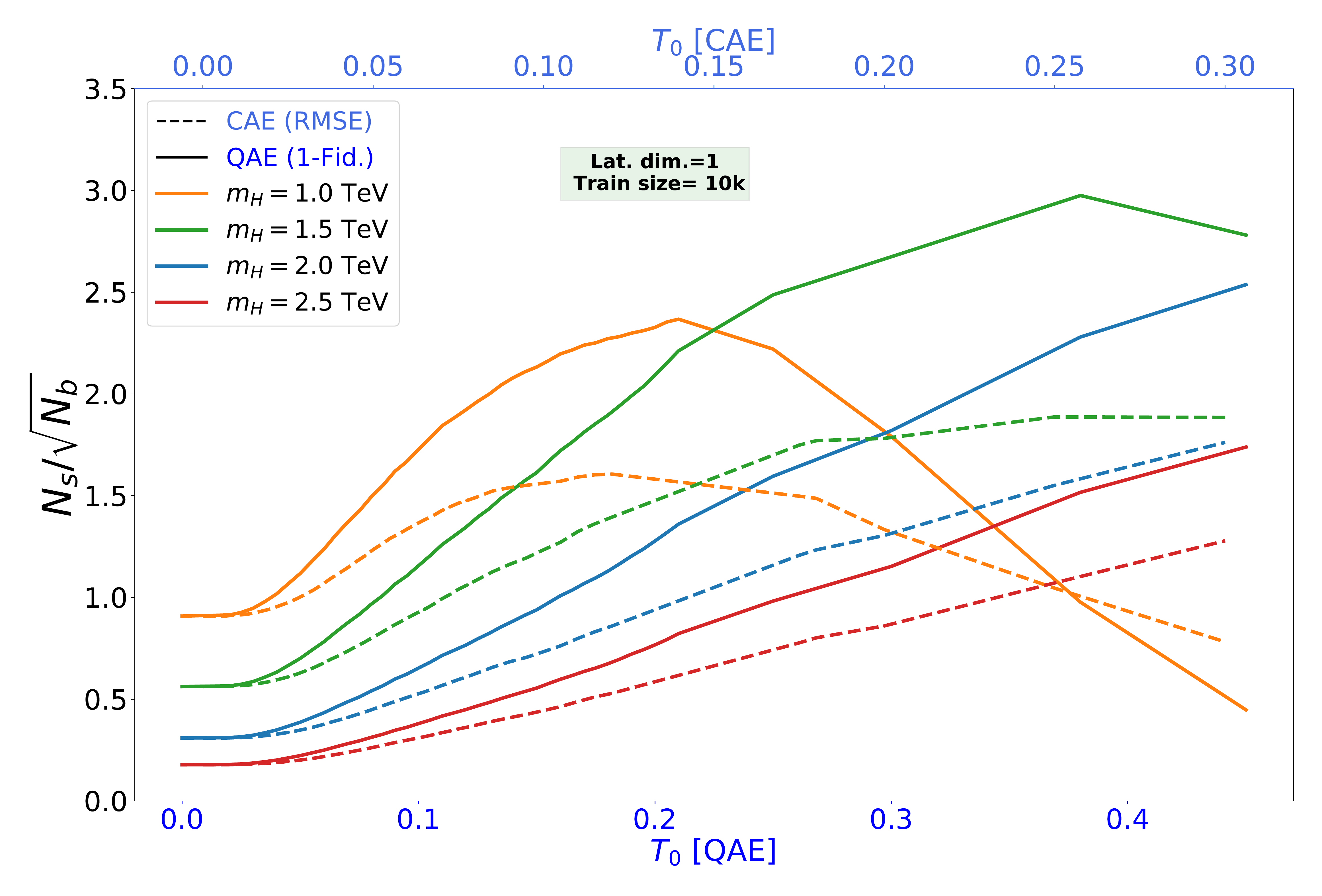}
	\caption{Significance as a function of the threshold $T_0$ on the fidelity and root-mean-square-error (RMSE) of the QAE and the CAE, respectively, for each of the signal scenarios and singular latent dimension trained on 10k samples. To keep the signal rich region on the right side for both, we have used $(1-\text{Fidelity})$ for the QAE. We fix the cross-section of all signals to 10 fb, and evaluate the yields at an integrated luminosity of 3000 fb$^{-1}$.}
	\label{fig:signi} 
\end{figure*}

 We compare the QAE and the CAE's performance for the four-dimensional input feature space. The metrics used in this presentation bear similarity to those used in a supervised framework. It also assumes that a randomly chosen event is equally likely to be either the background or the signal. This assumption is not sound in the context of LHC searches or in an anomaly detection technique since the background's cross-section is orders of magnitude larger than that of the signal. Nevertheless, they are handy when comparing different classifiers.

    For each value of $m_H$, we plot the Receiver-Operator-Characteristics (ROC) curve between the signal acceptance and the background rejection in figure~\ref{fig:roc}, for the networks trained with 10k samples. The ROC curve is obtained by evaluating the signal acceptance as a function of the background rejection since both are functions of the threshold $T_0$ applied on the loss function.  The signal acceptance $\epsilon_S\in[0,1]$ quantifies the fraction of accepted signal events when one puts a threshold $T_0$ on the variable $x$, while the background rejection $\bar{\epsilon}_B\in[0,1]$ measures the fraction of rejected background events for the threshold $T_0$.  An outline of how the ROC curve is obtained is given in Appendix~\ref{app:roc}.  The black dotted lines denote the performance of a random classifier with no knowledge of either the signal or the background, and the lines further away from it indicate better performance than those in its vicinity. The performance reduces with increasing latent dimensions for CAEs and QAEs, with the highest background rejection coming for a singular latent dimension. Comparing the QAEs and the CAEs (dotted vs solid lines for each colour), we find that QAEs perform better than CAEs consistently in all latent dimensions and the different values of $m_H$. This better performance may be a universal property of QAEs. However, as our analysis is a proof-of-concept, an in-depth exploration of the properties of QAEs in general and anomaly detection at colliders, in particular, is needed to affirm this observation.

\subsection{Anomaly detection}
\label{sec:tt_anom_det}
We now explore the performance of the autoencoders in a semi-realistic search scenario. When we scale the normalization of the signal and the background by their respective probability of occurrence, i.e. their respective cross-sections, we are essentially in an anomaly detection scenario since the background is orders of magnitude larger than the signal.  The performance of the autoencoders can then be quantified in terms of statistical significance as a function of the threshold applied on the loss. For the background, we scale the cross-section obtained from Madgraph by a global $k$ factor of 1.8~\cite{Czakon:2013goa}, while for all the signal masses, we fix a reference value of 10 fb.  The yield is then calculated as $$N_p=\epsilon_p\;\sigma_p\; L\; E_p(T_0)\quad,$$ where $\epsilon_p$ is the baseline selection efficiency, $\sigma_p$ the cross-section, and $E_p(T_0)$ the efficiency at a threshold $T_0$ of the loss distribution, for a process $p$, while $L$ is the integrated luminosity which we take to be 3000 fb$^{-1}$. 

\begin{figure*}[t]
	\centering
	\includegraphics[scale=0.5]{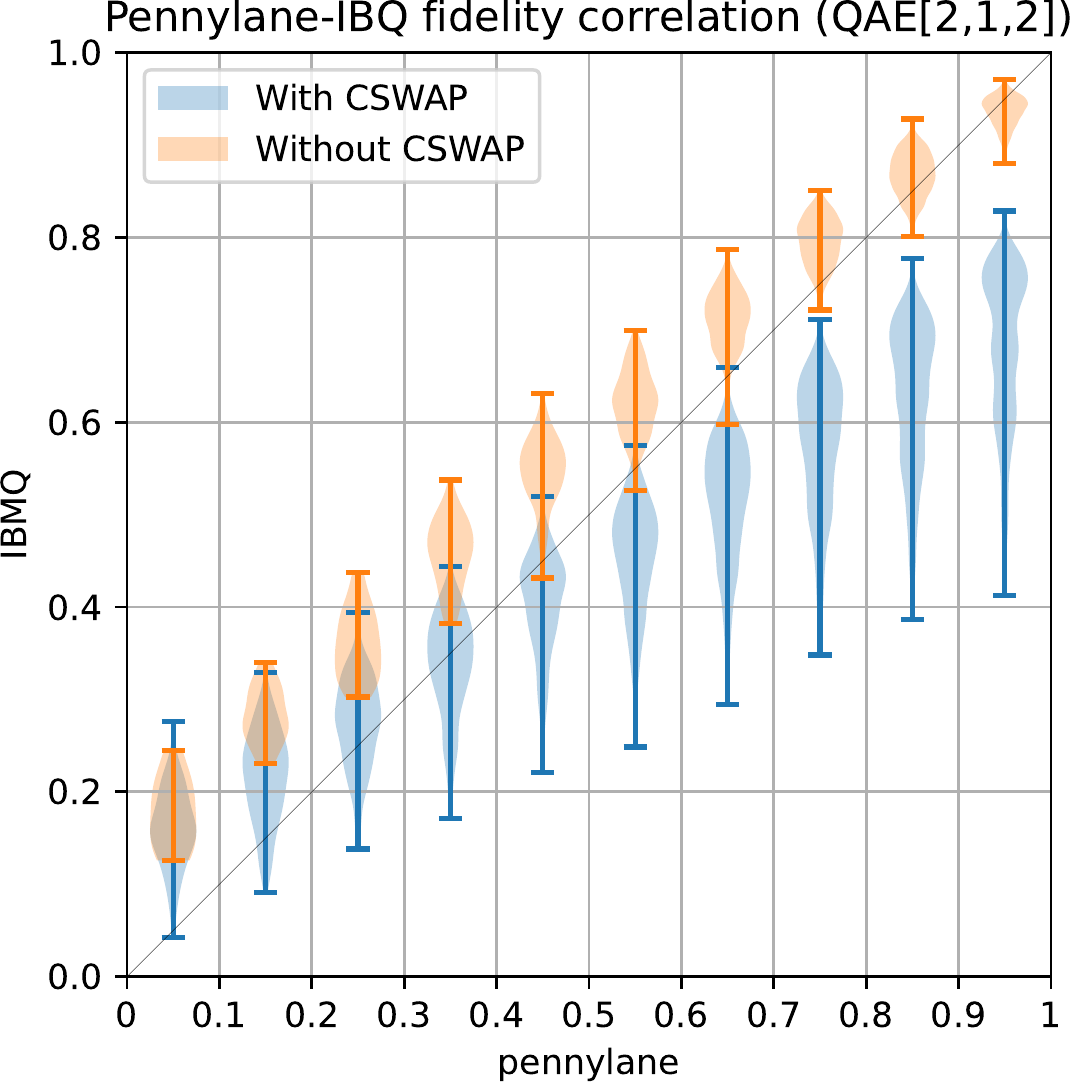}
	\hspace{1.0 cm}
	\includegraphics[scale=0.5]{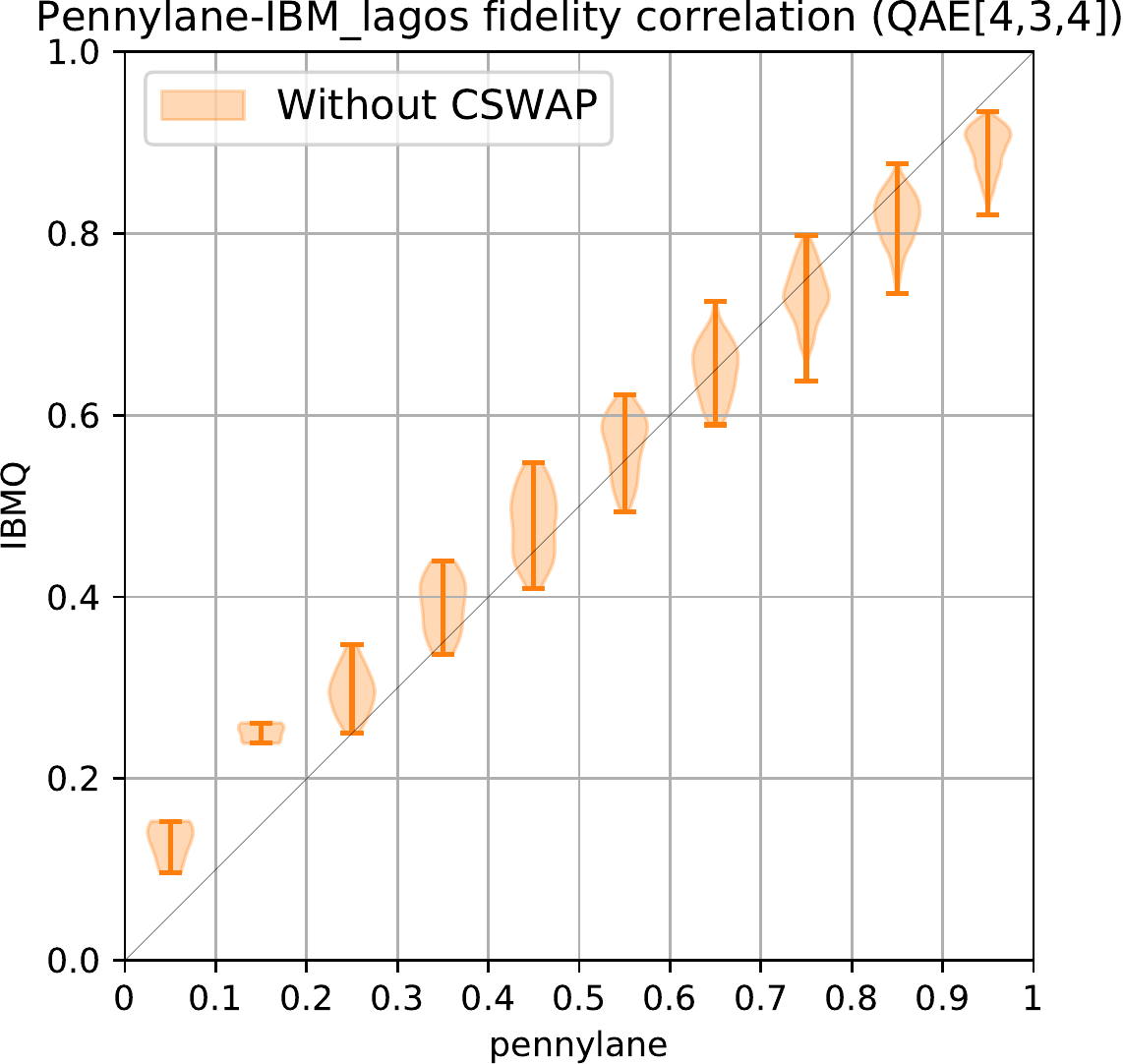}		
	\caption{The correlation between the fidelity values obtained by {\tt Pennylane} 
		and by the IBM-Q backends. On the left we show the comparison of a 2-1-2 QAE, where we directly measure the trash state (orange) and with a SWAP test employing a CSWAP gate. We find that the shallower implementation without the CSWAP gate has lesser decoherence effects, and hence better agreement with the simulation. The correlation with the direct measurement for the 4-3-4 case is shown on the right. 
	}
	\label{fig:ibm_qae} 
\end{figure*}

\begin{figure*}[t]
	\centering
	\includegraphics[scale=0.45]{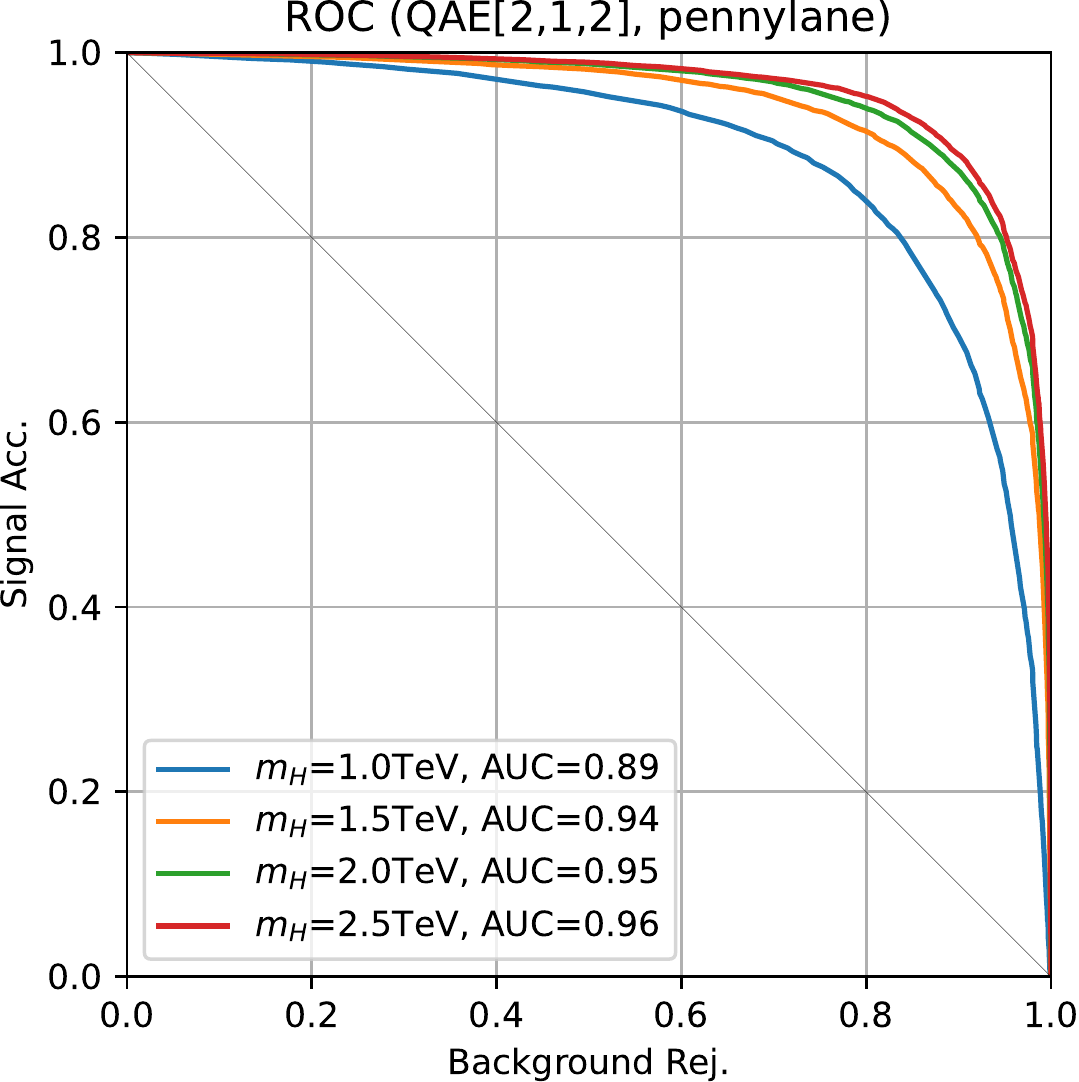}
	\includegraphics[scale=0.45]{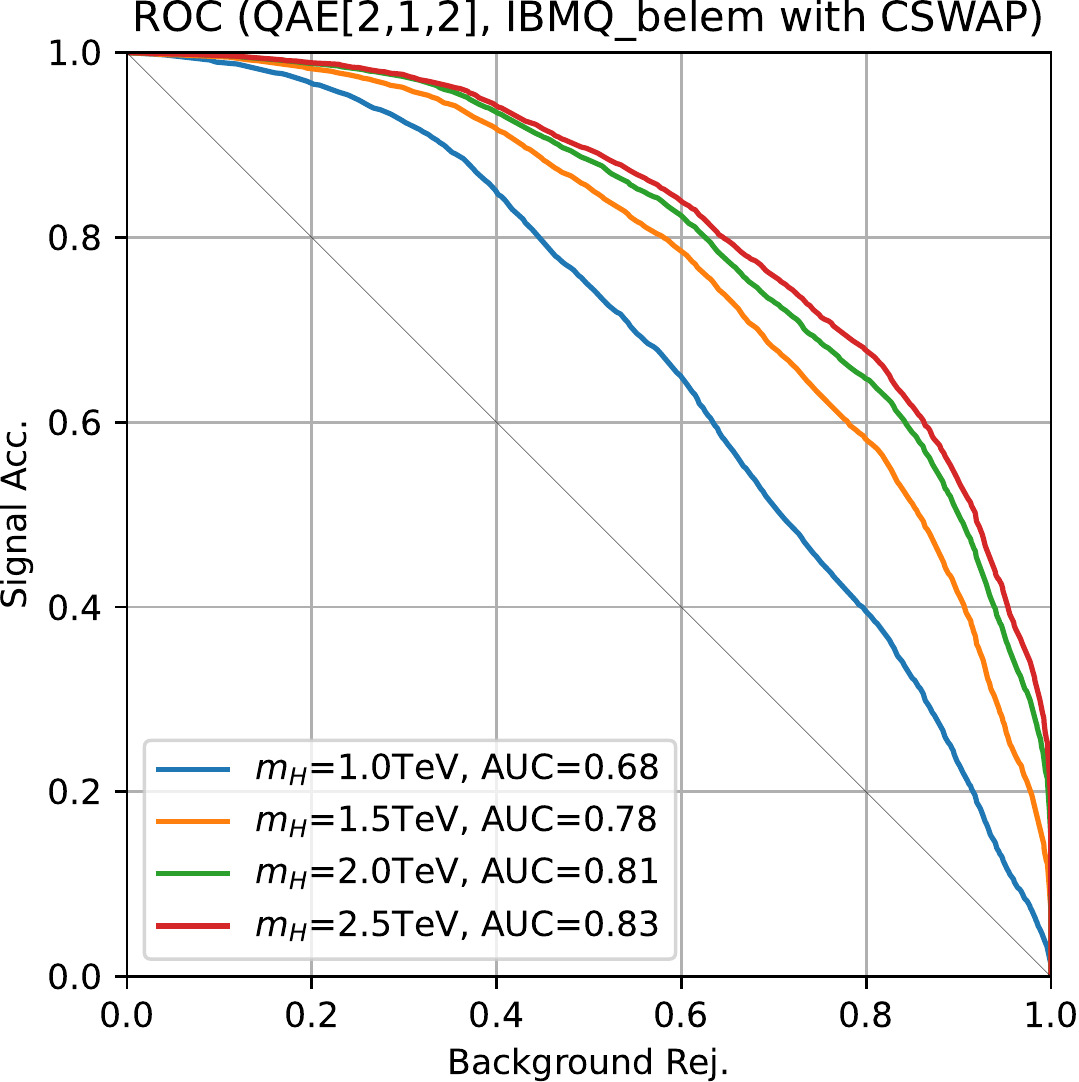}
	\includegraphics[scale=0.45]{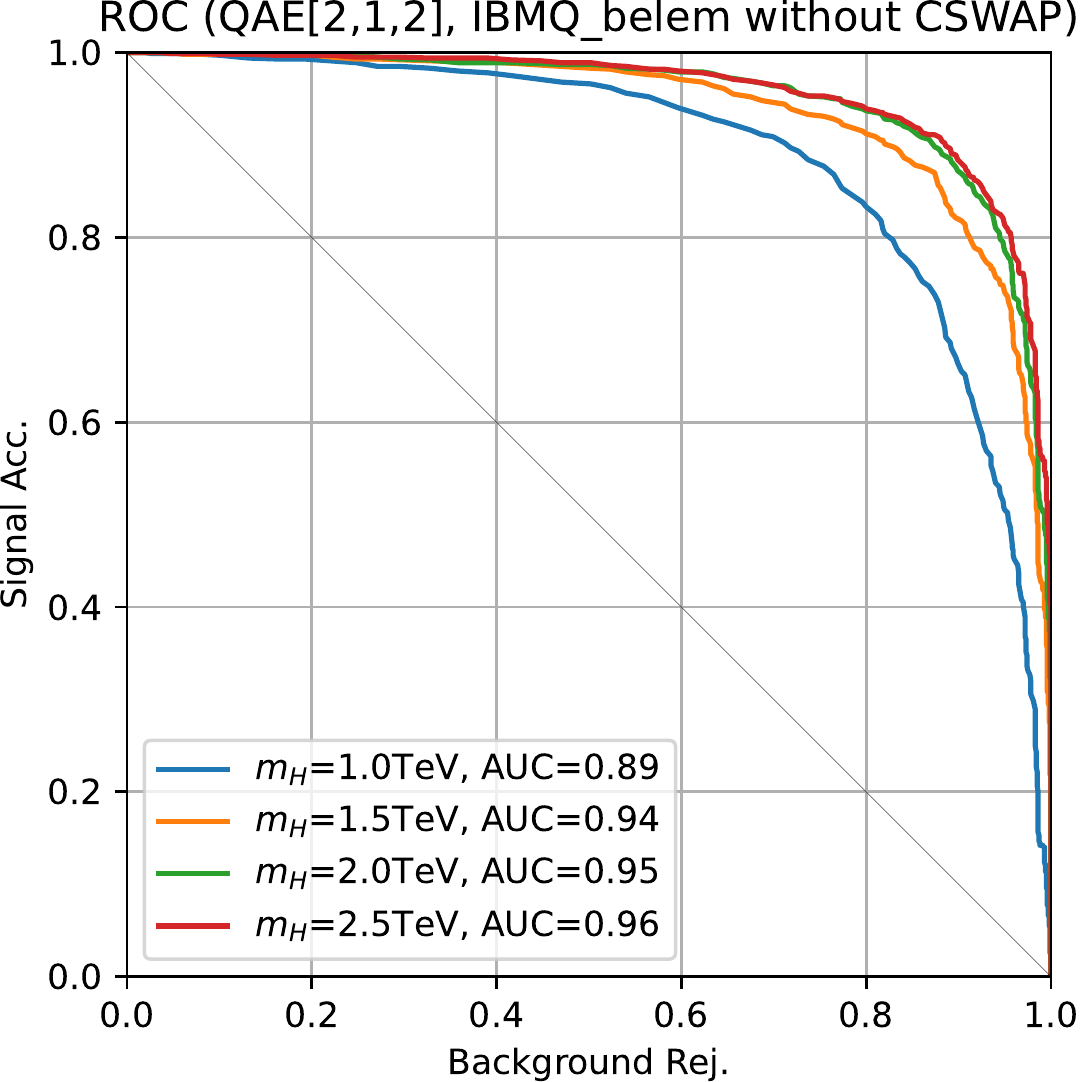}	
	\caption{ROC curves based on the fidelity distributions.  Those evaluated by the {\tt Pennylane} simulator (left panel),  
		by the quantum device IBM-Q belem backend with the SWAP test (central panel), and with the second qubit measurements (right panel)
		are shown.}
	\label{fig:ibm_roc} 
\end{figure*}

Since it is natural to use the best classifier in a search, we evaluate the significance of the autoencoders with one latent dimension, trained on 10k samples. We apply the threshold for the QAE and the CAE on the quantum trash state fidelity and the RMSE loss, respectively. We use $(1-\text{Fidelity})$ for the QAE to make the signal-rich regions same in both scenarios. RMSE loss is chosen over the cosine similarity since the former was found to have a higher performance.  The significance $N_S/\sqrt{N_B}$ for each of the signal masses as a function of the threshold $T_0$ is shown in figure~\ref{fig:signi}.  We fix the threshold range so that there are enough background test statistics in the least background like bin. Looking at the peak of the significance, we note that QAEs outperform CAEs, which is only natural from the preceding discussions.  However, an interesting development is the relative performance for the different masses. Even though the ROCs indicated higher discrimination with increasing mass, the significance increases for $m_H=1.0$ TeV to 1.5 TeV and decreases for higher masses. Since we have fixed a fiducial cross-section for each signal mass, it plays no role in this irregularity. The trend arises via an interplay between the higher discrimination by the autoencoder output and the decrease in baseline efficiency with increasing mass $m_H$. The decreasing selection efficiency is due to the isolation criteria of the jets and the leptons, which would be naturally boosted when we go to higher resonant masses $m_H$, thereby becoming more collimated.

\subsection{Benchmarking on a quantum device}
\label{sec:tt_ibmq}
We now compare the performance of the quantum simulator and the actual quantum hardware. 
Since there is a limitation on the available number of qubits, 
we limit the feature space in two dimension, which consists of $\{p^{b_1}_T,p^{l_1}_T\}$. 
For our QAE setup, in addition to the two qubits for embedding the input features, 
one qubit for the reference state and another ancilliary qubit for the SWAP test are needed. 
We use the simpler version of the quantum circuit shown in figure~\ref{fig:q_circuit}, which is implemented and trained using {\tt PennyLane}. To compare the performance, 
we use the same circuit with the same optimized parameters both for 
{\tt PennyLane} and for the IBM-Q belem backend. Accessing the IBM hardware was done through {\tt Qiskit}.

In figure~\ref{fig:ibm_qae}, we show the fidelity distributions for the background and the signal samples
for our QAE circuit with the optimized circuit parameters computed by the simulator 
in {\tt Pennylane} and in the actual quantum device of IBM-Q belem backend. The plot shows the shape of the distribution (denoted by the width of the shaded region) in the y-axis for each bins of size 0.1 in the x-axis (plotted at each bin center). The lines at each ending denote the range of the data of the y-axis.
Since IBM-Q does not have a shallow implementation of the CSWAP operation, 
the fidelity distributions are smeared toward 0.5, and it is especially worse around 1.
One of the advantages of using the SWAP test is to reduce the number of qubits for the evaluations of the fidelity 
during the optimization process. For example, to check the performance of the current circuit, 
directly measuring the fidelity between the reference state and the output for the second qubit
would be enough. It can be achieved by the simple Pauli $z$ measurements.
The correlation of the fidelities obtained by {\tt Pennylane} and by IBM-Q belem, based on the SWAP test 
and on the Pauli $z$ measurement are shown 
in the right panel as the violin plots, in blue and in orange, respectively.
The correlation is better for the Pauli $z$ measurements for the same circuit part with the identical input parameters.
It suggests that the decoherence effects from a deeper circuit obscure the performance.

In figure~\ref{fig:ibm_roc}, we show the ROC curves based on the 
fidelity distributions for the background and the signal samples
evaluated by {\tt Pennylane} simulator in the left panel. 
The central panel shows the ROC curves 
based on the fidelities evaluated by the SWAP test, while the right panel shows those by the second qubit Pauli $z$ measurements, 
for the same IBM-Q device of belem backend. 
As one can see, 
the performances based on the Pauli $z$ measurements on the IBM-Q device follow those obtained by the {\tt Pennylane} simulator.
The AUCs for them are also essentially the same. 
Thus, the deficit in the performance with the SWAP test is due to the too deep circuit realization 
for the CSWAP operation in the IBM-Q device. 
Therefore, the realisation of a CSWAP operation with a shallow circuit is necessary.

To check the efficacy of quantum hardware for the four input QAE, we evaluate
the trash state fidelity of a QAE with four-dimensional input features. Due to hardware limitations discussed above, we estimate it without the SWAP test for a single trash qubit giving us a three-dimensional latent representation. The correlation between the {\tt Pennylane} evaluated fidelity and the output from IBM-Q lagos, shown in figure~\ref{fig:ibm_qae}, displays a good agreement between the simulation and the hardware.
\subsection{Comparative training efficiency and performance for $pp\to Z(\nu\bar{\nu})jj$ background}
\label{sec:z_inv_comp}
\begin{figure*}[t]
	\centering
	\includegraphics[scale=0.22]{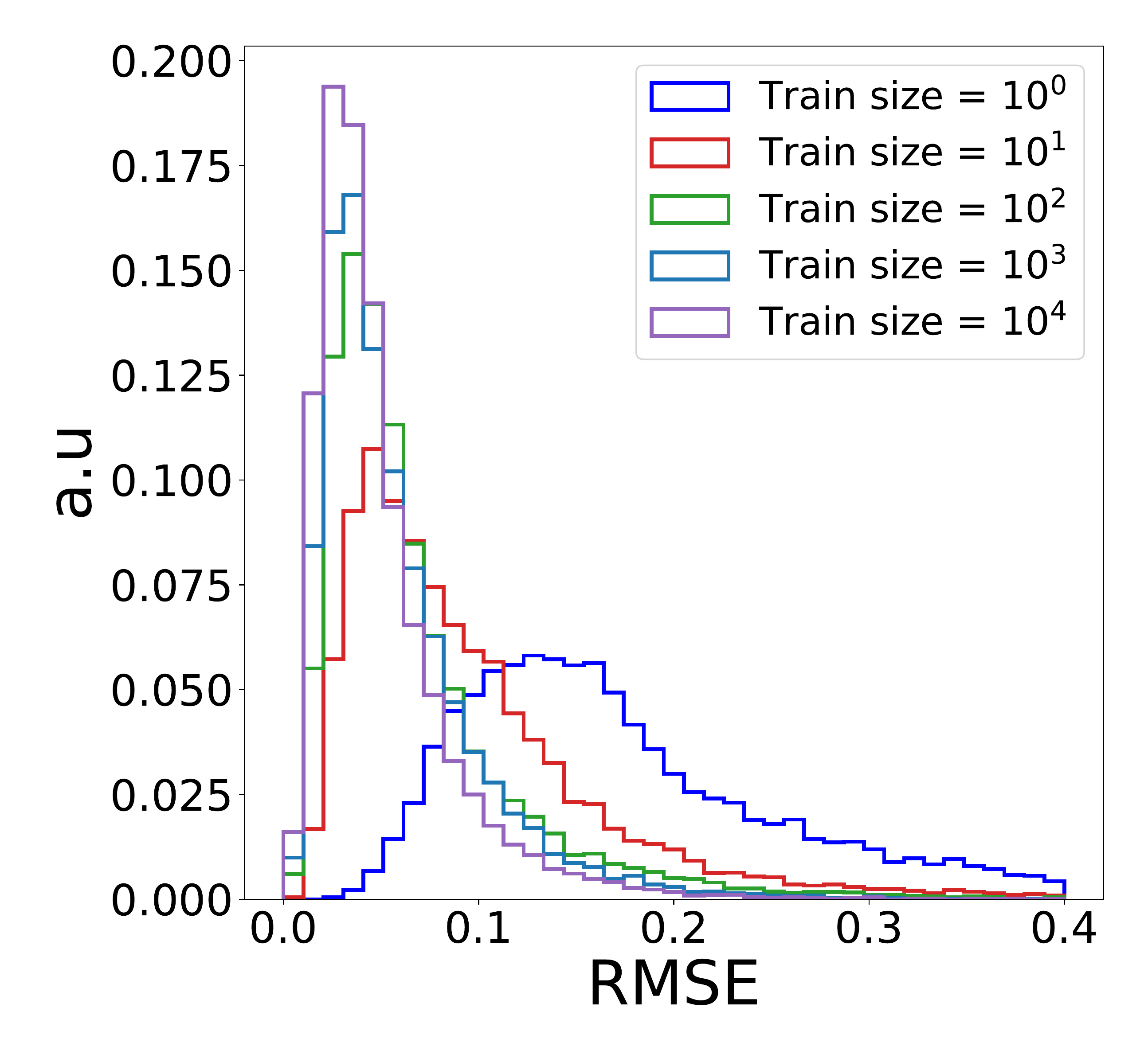}
	\includegraphics[scale=0.22]{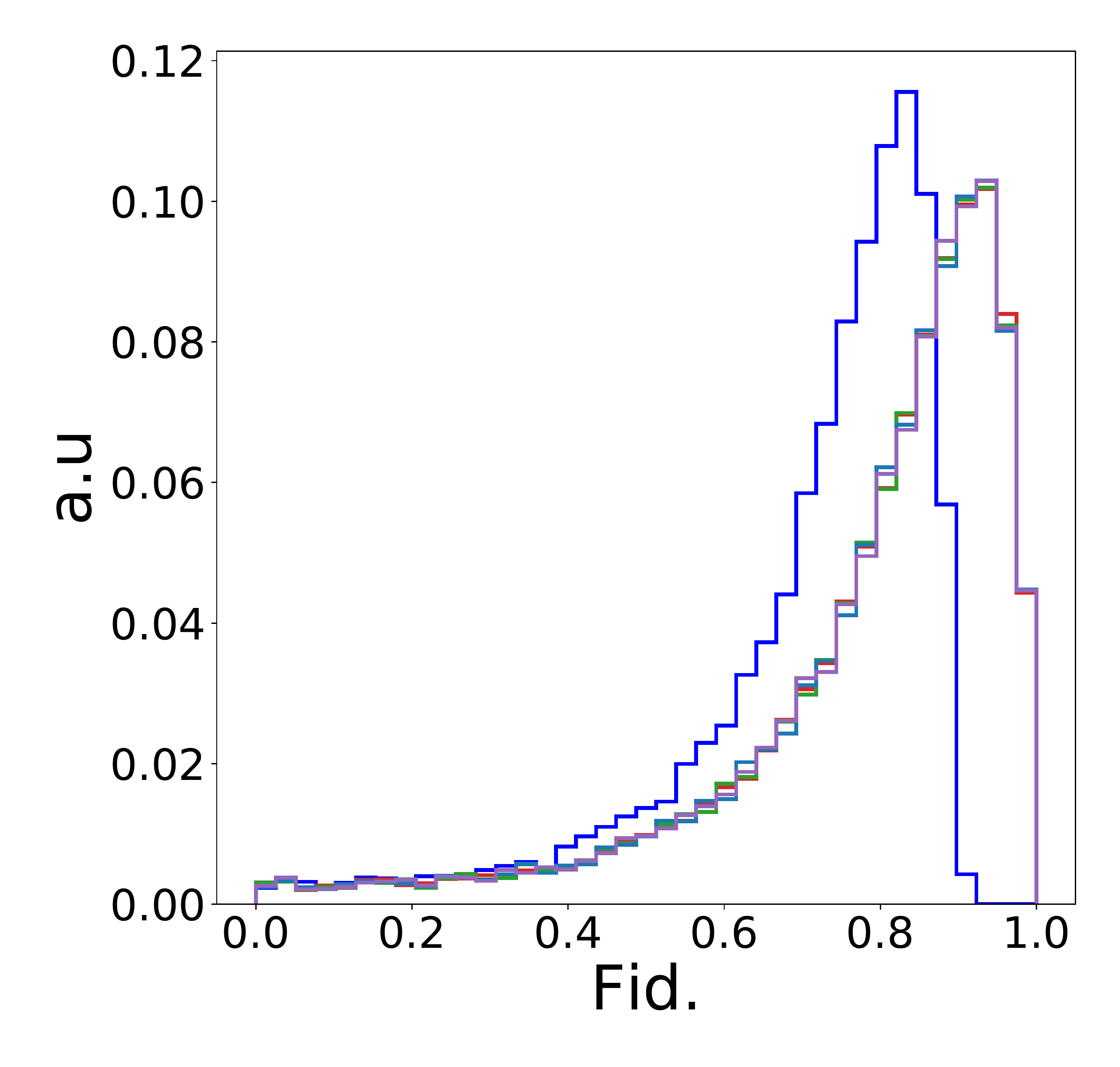}
	\includegraphics[scale=0.22]{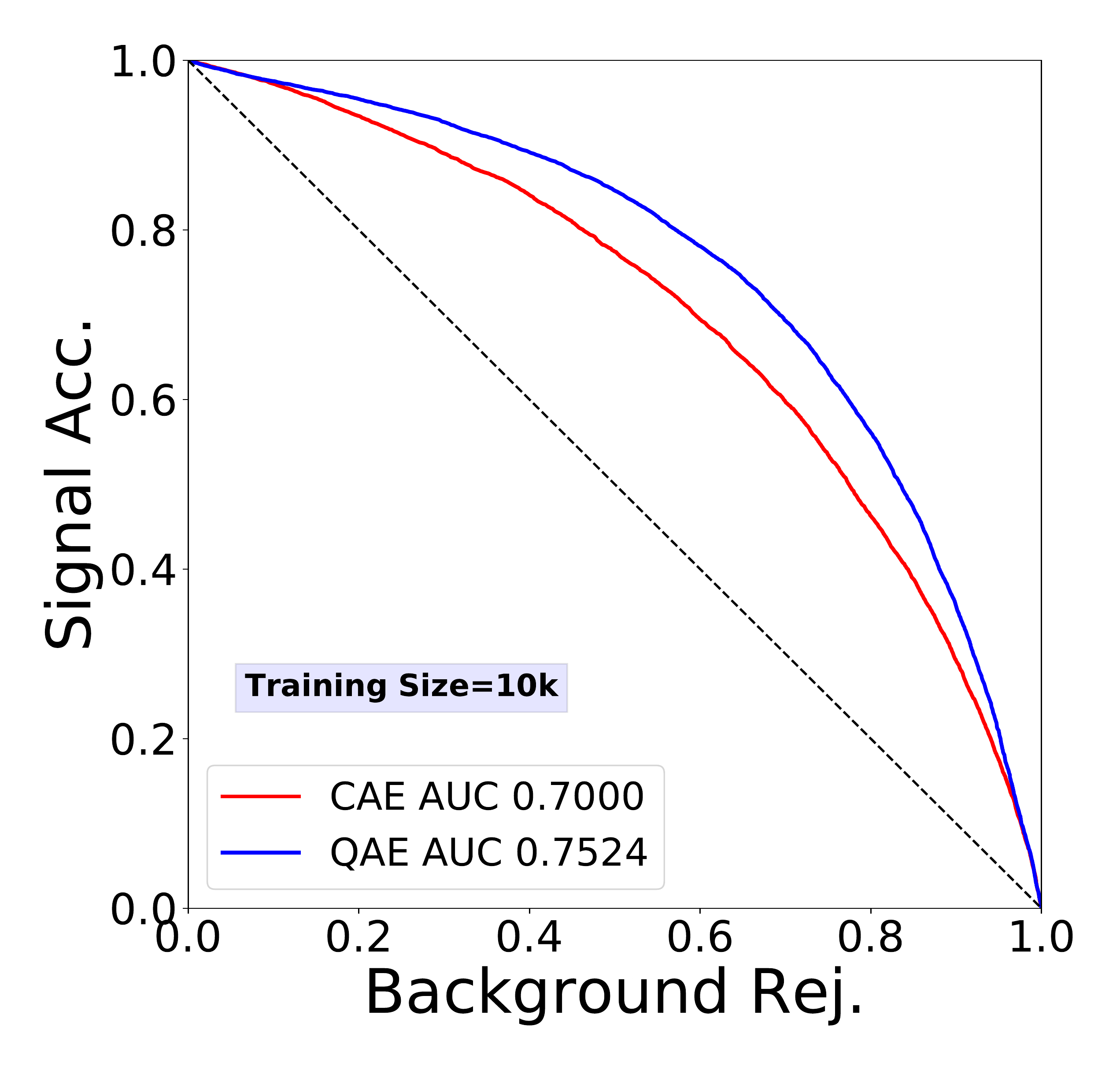}
	\caption{The test distribution of the invisible $Z$ background scenario for different training sizes of a CAE (left) and a QAE (center) for a two-dimensional latent representation, and their respective ROC curve (right) for the training done with 10k events. Similar to the previous case, the QAE has converged with much smaller datasets than the CAE. Moreover, the QAE performs relatively better than the CAE for the particular signal.}
	\label{fig:inv_sig_bg_sep} 
\end{figure*}

	We have seen that a QAE trains efficiently and performs better than a CAE in a hypothetical resonant signal scenario. To gauge how these important behaviours carry over to a different process, we study the training size dependence and performance of a QAE and CAE for an invisible background (and signal), detailed in the last paragraph of Section~\ref{sec:data_sim} for a two-dimensional latent space. Note that all the results for the CAE are for the best model chosen after a hyperparameter scan described in Appendix~\ref{app:hyp_scan}.
	
The loss distribution of the test dataset for the background for different sizes of training data and their ROC curve for the case of 10k training samples are shown in figure~\ref{fig:inv_sig_bg_sep}. The characteristics are similar to the previous scenario, giving further evidence that the training efficiency of the QAE is not limited to a specific kind of process. Moreover, from the ROC and the AUC value, we see that the QAE also performs better than the CAE. This superior performance is particularly noteworthy given that the CAE's hyperparameters has been chosen after a hyperspace scan restricted to a fixed width and depth.

\section{Conclusion}
  \label{sec:conc}
The lack of evidence for new interactions and particles at the Large Hadron Collider has motivated the high-energy physics community to explore model-agnostic data-driven approaches to search for new physics. Machine-learning anomaly detection methods, such as autoencoders, have shown to be a powerful and flexible tool to search for outliers in data. Autoencoders learn the kinematic features of the background data by training the network to minimise the reconstruction error between input features and neural network output. As the kinematic characteristics of the signal are different to the background, the reconstruction error for the signal is expected to be larger, allowing signal events to be identified as anomalous. 

Although quantum architecture capable of processing huge volumes of data is not yet feasible, noisy-intermediate scale devices could have very real applications at the Large Hadron Collider in the near future. With the origin of the collisions being quantum-mechanical, a quantum autoencoder could, in principle, learn quantum correlations in the data that a bit based autoencoder fails to see. We have shown that quantum-autoencoders based on variational quantum circuits have potential applications as anomaly detectors at the Large Hadron Collider. Our analysis shows that for the scenario we consider, i.e. the same set of input variables, quantum autoencoders outperform dense classical autoencoders based on artificial neural networks, asserting that quantum autoencoders can indeed go beyond their classical counterparts. They are very judicious with data and converge with very small training samples. This independence opens up the possibility of training quantum autoencoders on small control samples, thereby opening up data-driven approaches to inherently rare processes.

\section*{Acknowledgements}
We acknowledge the use of IBM-Q in this work and thank Simon Williams for technical support. M.S. is supported by the STFC under grant ST/P001246/1. VSN is supported by the Physical Research Laboratory (PRL), Department of Space, Government of India. We thank Anupam Ghosh, Partha Konar, and Sudipta Show for valuable discussions on quantum machine learning. 
MT is supported, in part, by 
the Grant-in-Aid for Scientific Research\,C, No.~18K03611, 
the Grant-in-Aid on Innovative Areas, the Ministry of Education, Culture, Sports, Science and Technology, No. 16H06492, 
and by the JSPS KAKENHI Grant No.~20H00160.

\appendix
\section{Quantum Gradient Descent}
\label{app:qgd}
	
We discuss the basic idea behind quantum gradient descent~\cite{Stokes2020quantumnatural} in this appendix. The general idea is to make the optimisation procedure aware of the underlying quantum geometry of the weight space. Denoting any generic weight vector by $\Theta$, we have the vanilla gradient descent update as, 
\begin{equation}
\label{eq:sgd} 
\Theta_{i+1}=\Theta_i-\gamma\;\nabla_\Theta \;L(\Theta)\quad,
\end{equation}where $L$ is a well-behaved loss function. This expression implicitly assumes that $l_2$ distances correctly describe the underlying geometry of the weight space, placing all directions in the weight space on an equal footing. In reality, however, the geometry of the weight space can be much more complicated, and such a straightforward update rule may not converge to the optimal point. Therefore, to have an idea of the underlying geometry, we modify eq.~\ref{eq:sgd} with the metric tensor $\mathbf{G}$, 
\begin{equation}
\label{eq:ngd} 
\Theta_{i+1}=\Theta_i-\gamma\;\mathbf{G}^{-1}(\Theta_i)\;\left(\nabla_\Theta L(\Theta)\right)_{\Theta=\Theta_i}\quad,
\end{equation} to get the Natural Gradient descent~\cite{10.1162/089976698300017746}. Note that Natural Gradient descent gives the usual gradient descent (eq.~\ref{eq:sgd}) for a Euclidean metric $\mathbf{G}=\mathbf{I}$. Due to the extremely large parameter space, it is computationally prohibitive to put metric-restrained optimisation in deep neural networks, which is not the case for currently used variational quantum circuits. 
 The natural metric on complex projective Hilbert Spaces (the space containing physical quantum states) is the Fubiny-Study metric~\cite{fubini,Study1905},
 \begin{equation}
 g_{ij}=\text{Re}[\;\braket{\partial_i\phi_0}{\partial_j\phi_0}-\braket{\partial_i\phi_0}{\phi_0}\;\braket{\phi_0}{\partial_j\phi_0}\;]\quad.
 \end{equation}
 Here, $\ket{\partial_i\phi_0}=\frac{\partial \ket{\phi_0}}{\partial \theta_i}$, with $\theta_i$, a component of the weight vector $\Theta$ and $\ket{\phi_0}$, a state in the Hilbert space. The inverse of the metric is evaluated in {\tt Pennylane} using the Moore-Penrose pseudo inverse~\cite{bams/1183425340,penrose_1955} $$g^+ = (g^T g)^{−1} g^T\quad,$$ which is well-behaved even when $\det g=0$ and is numerically equal to the inverse when it exists. 
\section{Defining the Receiver-Operator-Characteristics curve}
\label{app:roc} 

In this appendix, we outline the procedure of obtaining the ROC curve from the normalised probability distribution of a signal $p_S(x)$, and background $p_B(x)$. By normalised, we mean $\int_{-\infty}^{\infty}\;dx\;p_S(x)=\int_{-\infty}^{\infty}\;dx\;p_B(x)=1$. 
	The signal acceptance $\epsilon_S=f_S(T_0)$ and the background rejection $\bar{\epsilon}_B=\bar{f}_B(T_0)$ are defined as
	
	\begin{equation*}
	f_S(T_0)=\int_{-\infty}^{T_0}\;dx\; p_S(x)\quad,\quad \bar{f}_B(T_0)=\int_{T_0}^{\infty}\;dx\; p_B(x)\quad,
	\end{equation*}
	where we have assumed that the signal rich regions are on the lower side of the variable $x$. The ROC curve is then obtained by expressing the signal acceptance as a function of the background rejection as
	$$T_0=\bar{f}^{-1}_B(\bar{\epsilon}_B)\implies \epsilon_S=f_S(T_0)=f_S(\bar{f}^{-1}_B(\bar{\epsilon}_B))\quad.$$ The ROC curve therefore shows the function $\epsilon_S(\bar{\epsilon}_B)$ without any reference to the threshold $T_0$, which is implicitly assumed in the evaluation of the dependent ($\epsilon_S)$ and the independent ($\bar{\epsilon}_B$) quantities. The variable $x$ can be any physical observable or the output of a neural network model. For the studies conducted here, it is the RMSE loss for the CAE, and the fidelity for the QAE.

\section{Details of hyperparameter scan}
\label{app:hyp_scan}

\begin{table}[h]
	\begin{tabular}{|c|l|l|l|}
		\hline
		Sl. no. &Hyper Parameter     & Value Space              & Best value  \\ \hline
		1.&Activation function & tanh, ReLu, Sigmoid, Linear & ReLu     \\ [0.1cm]
		2.&L1 Regularisation   & 0,0.1,0.01,0.001,0.0001 &0\\ [0.1cm]
		3.&L2 Regularisation   & 0,0.1,0.01,0.001,0.0001 & 0   \\ [0.1cm]
		4.&Dropout& 0,0.1,0.2,0.3&0 \\ [0.1cm]
		5.&Learning Rate&0.01,0.001,0.0003&0.0003\\ [0.1cm]
		6.& Batch Size&32,64,128,256,512,1024&64\\ [0.1cm]
		 \hline
	\end{tabular}
	\caption{The table shows the different values of the hyperparameters and their best values after the scan.}
	\label{tab:hyper} 
\end{table}

	The details of the hyperparameter scan of classical autoencoder with six-dimensional inputs and outputs are given in this appendix. We use the {\tt RandomSearch} algorithm implemented in {\tt KerasTuner}~\cite{omalley2019kerastuner} for the scan. The number of nodes in the hidden layers of the encoder is kept fixed to 20, 15, and 10. With a (fixed) two-dimensional latent space, we use a symmetric decoder setup. Once the skeleton of the architecture is fixed, we scan over the activation function of the layers,  L1 regularisation and L2 regularisation of the weights, the dropout value between two successive layers, and the training's learning rate and batch size. Their respective values along with the best one chosen for the final training are given in table~\ref{tab:hyper}. The best value of the hyperparameters are from thousand trials trained for hundred epochs, and the training is terminated if the validation loss does not improve for ten epochs (implemented as the {\tt EarlyStopping} callback during training).
	
	We do not vary the width or the depth to compare the capabilities of CAEs with at least some degree of comparability to the simple QAE used in the study. Increasing the width and depth will undoubtedly increase the expressive power of a CAE, which is not the objective of the current study. Networks like Convolutional or graph autoencoders acting on low-level high dimensional data will undoubtedly perform better than currently executable QAEs. However, existing quantum resources cannot process such high dimensional data. 
	
\bibliographystyle{apsrev4-1}
\bibliography{ref.bib}
	
\end{document}